\documentclass[journal]{IEEEtran}
\IEEEoverridecommandlockouts
\usepackage{setspace}               
\usepackage{stfloats}
\usepackage{cite}
\usepackage{amsmath,amssymb,amsfonts}
\usepackage{upgreek}
\usepackage{graphicx}
\usepackage{subfigure}
\usepackage{textcomp}
\usepackage{xcolor}
\usepackage{algorithm}
\usepackage[noend]{algpseudocode}
\usepackage{amsmath}
\usepackage[caption=false,font=normalsize,labelfont=sf,textfont=sf]{subfig}
\usepackage[linewidth=1pt]{mdframed}
\usepackage[mathscr]{eucal}
\usepackage{balance}
\usepackage{epstopdf}
\usepackage{cases}
\usepackage{xcolor}
\usepackage{bbding}
\usepackage{cleveref}
\usepackage{multicol}       
\usepackage{multirow}       
\usepackage{array}          
\usepackage{makecell}
\usepackage{flushend}
\usepackage{booktabs}
\definecolor{crimson}{RGB}{192,0,0}         
\definecolor{navy}{RGB}{47,85,151}         
\usepackage{colortbl}

\def\BibTeX{{\rm B\kern-.05em{\sc i\kern-.025em b}\kern-.08em
    T\kern-.1667em\lower.7ex\hbox{E}\kern-.125emX}}
\renewcommand{\arraystretch}{1.5}

\begin{document}
\title{Cooperative Multi-Target Positioning for Cell-Free Massive MIMO with Multi-Agent Reinforcement Learning}
\author{{Ziheng~Liu,~\IEEEmembership{Student Member,~IEEE}, Jiayi~Zhang,~\IEEEmembership{Senior Member,~IEEE}, Enyu~Shi,~\IEEEmembership{Student Member,~IEEE}, Yiyang~Zhu, Derrick~Wing~Kwan~Ng,~\IEEEmembership{Fellow,~IEEE}, and Bo~Ai,~\IEEEmembership{Fellow,~IEEE}}
\thanks{This work was supported by the Fundamental Research Funds for the Central Universities under Grant No. 2024YJS138, in part by National Natural Science Foundation of China under Grant 62471027, in part by National Natural Science Foundation of China under Grant 62221001, in part by Natural Science Foundation of Jiangsu Province, Major Project under Grant BK20212002, in part by ZTE Industry-University-Institute Cooperation Funds under Grant No. IA20240709018, and in part by ZTE Industry-University-Institute Cooperation Funds under Grant No. IA20240319002. Part of this article has been accepted at IEEE SPAWC 2024 \cite{[43]}. \itshape(Corresponding author: Jiayi Zhang.)\upshape}
\thanks{Z. Liu, J. Zhang, E. Shi, Y. Zhu, and B. Ai are with the School of Electronic and Information Engineering and also with the Frontiers Science Center for Smart High-Speed Railway System, Beijing Jiaotong University, Beijing 100044, China (e-mail: \{23111013, zhangjiayi, 21111047, 21251058, boai\}@bjtu.edu.cn).}
\thanks{D. W. K. Ng is with the School of Electrical Engineering and Telecommunications, University of New South Wales, NSW 2052, Australia (e-mail: w.k.ng@unsw.edu.au).}}
\maketitle
\vspace{-1.75cm}
\begin{abstract}
Cell-free massive multiple-input multiple-output (mMIMO) is a promising technology to empower next-generation mobile communication networks.
In this paper, to address the computational complexity associated with conventional fingerprint positioning, we consider a novel cooperative positioning architecture that involves certain relevant access points (APs) to establish positioning similarity coefficients. Then, we propose an innovative joint positioning and correction framework employing multi-agent reinforcement learning (MARL) to tackle the challenges of high-dimensional sophisticated signal processing, which mainly leverages on the received signal strength information for preliminary positioning, supplemented by the angle of arrival information to refine the initial position estimation. Moreover, to mitigate the bias effects originating from remote APs, we design a cooperative weighted K-nearest neighbor (Co-WKNN)-based estimation scheme to select APs with a high correlation to participate in user positioning.
In the numerical results, we present comparisons of various user positioning schemes, which reveal that the proposed MARL-based positioning scheme with Co-WKNN can effectively improve positioning performance.
It is important to note that the cooperative positioning architecture is a critical element in striking a balance between positioning performance and computational complexity.
\end{abstract}
\begin{IEEEkeywords}
Cell-free massive MIMO, cooperative WKNN, multi-agent reinforcement learning, user positioning.
\end{IEEEkeywords}

\IEEEpeerreviewmaketitle
\section{Introduction}
The goal of next-generation mobile communication networks, including beyond fifth-generation (B5G) and sixth-generation (6G) networks, is to support the exceptional popularity of smart devices and the enormous growth of mobile data traffic, thereby realizing the vision of the Internet-of-Everything (IoE) \cite{[1],[41]}. Among the numerous 6G technologies \cite{[20],[22],[45]}, cell-free massive multiple-input multiple-output (mMIMO) technique holds a prominent place due to its ability to provide high spectral efficiency (SE), ensure reliable massive access, and suppress multi-user interference. In essence, cell-free mMIMO is a synergistic integration paradigm that harnesses the advantages of two cornerstone technologies, combining mMIMO with favorable propagation characteristics and network MIMO with uniform user performance \cite{[23],[24],[44]}.

Compared with conventional cellular network and mMIMO technologies \cite{[21]}, cell-free mMIMO technology represents an advanced network architecture consisting of a large number of collaborative access points (APs), which eliminates the original concept of a cell \cite{[22],[23]}. In this architecture, all APs coherently serve all user equipments (UEs) without any cell boundaries, performing spatial multiplexing using the same time-frequency resources \cite{[3],[29]}. As such, cell-free mMIMO is a significant leap forward in alleviating inter-cell interference in mMIMO technology, eliminating the fundamental limitations that inter-cell interference imposes on the performance of dense cellular networks \cite{[25],[26]}.
\subsection{Related Work}
With the widespread deployment of positioning-based applications in mobile communication networks, the design of accurate positioning technology has received significant attention in both industry and academia \cite{[15],[27],[35]}. Although conventional global positioning systems can provide real-time positioning for outdoor mobile users with an accuracy of a few meters, their positioning performance significantly declines in urban areas due to obstructions caused by buildings, cars, and pedestrians.

Recently, the strategy utilizing the rich information provided by multipath wireless propagation channels to achieve user positioning has garnered considerable momentum \cite{[7],[13],[14],[16],[18],[17]}. Numerous efficient positioning schemes have been proposed, such as the geometry positioning \cite{[28]}, fingerprint positioning \cite{[7],[8],[9],[10],[11],[16]}, iterative positioning \cite{[12],[14]}, and machine learning-based positioning \cite{[8],[15],[19]}. Specifically, geometry positioning \cite{[28]} belongs to a trilateral or triangular positioning scheme that relies heavily on the signals transmitted between APs and UEs, such as received signal strength (RSS) and time of arrival (TOA) under trilateral positioning, and the angle of arrival (AOA) under triangular positioning. However, in complex urban environments, the performance of geometric positioning is significantly reduced due to non-line-of-sight (NLoS) propagation phenomena, such as building occlusion.
In contrast, fingerprint positioning alleviates the challenge of NLoS propagation by deploying high-density and uniformly distributed reference points in proximity to users within a broad area, making it closer to the LoS propagation scenario. It mainly consists of two stages: offline and online. The former establishes a fingerprint dataset through the deployed reference points, while the latter transforms the positioning problem into an identification problem and achieves user positioning by searching the fingerprint dataset \cite{[7],[8],[9],[10],[11],[16]}. In particular, the most commonly adopted similarity criteria in fingerprint positioning are RSS and AOA \cite{[7],[8],[9]}, where RSS mainly extracts distance information exploiting large-scale fading information, while AOA mainly extracts angle information using channel state information.
\begin{table*}[t]
  \centering
  \fontsize{9}{10}\selectfont
  \caption{Comparison of Relevant Research With This Paper.}
  \label{Paper_comparison}
    \begin{tabular}{ !{\vrule width1.2pt}  m{1.5 cm}<{\centering} !{\vrule width1.2pt}   m{1.5 cm}<{\centering} !{\vrule width1.2pt}  m{1.5 cm}<{\centering} !{\vrule width1.2pt} m{1.8 cm}<{\centering} !{\vrule width1.2pt} m{1.4 cm}<{\centering}  !{\vrule width1.2pt} m{1.5 cm}<{\centering} !{\vrule width1.2pt} m{1.85 cm}<{\centering} !{\vrule width1.2pt} m{1.1 cm}<{\centering} !{\vrule width1.2pt}m{1.7 cm}<{\centering} !{\vrule width1.2pt}}
    \Xhline{1.2pt}
        \rowcolor{gray!30} \bf Ref. & \bf Cell-free mMIMO & \bf Positioning& \bf Cooperative Architecture & \bf Learning-based &  \bf Scalability  &  \bf Decentralized & \bf MARL & \bf Multi-layer Correction \cr
    \Xhline{1.2pt}
         \cite{[9],[10]}   & \makecell[c]{\Checkmark} & \makecell[c]{\Checkmark}& \makecell[c]{\XSolidBrush} & \makecell[c]{\XSolidBrush} & \makecell[c]{\XSolidBrush} & \makecell[c]{\XSolidBrush} & \makecell[c]{\XSolidBrush} & \makecell[c]{\XSolidBrush}\cr\hline
          \cite{[19]} & \makecell[c]{\XSolidBrush} & \makecell[c]{\Checkmark}& \makecell[c]{\Checkmark} & \makecell[c]{\Checkmark} & \makecell[c]{\Checkmark} & \makecell[c]{\XSolidBrush} & \makecell[c]{\XSolidBrush} & \makecell[c]{\XSolidBrush} \cr\hline
          \cite{[37]} & \makecell[c]{\XSolidBrush} & \makecell[c]{\Checkmark}& \makecell[c]{\Checkmark} & \makecell[c]{\Checkmark} & \makecell[c]{\Checkmark} & \makecell[c]{\Checkmark} & \makecell[c]{\XSolidBrush} & \makecell[c]{\XSolidBrush} \cr\hline
          \cite{[8]} & \makecell[c]{\Checkmark} & \makecell[c]{\Checkmark}& \makecell[c]{\XSolidBrush} & \makecell[c]{\Checkmark} & \makecell[c]{\XSolidBrush} & \makecell[c]{\XSolidBrush} & \makecell[c]{\XSolidBrush} & \makecell[c]{\XSolidBrush} \cr\hline
          \cite{[5],[31]} & \makecell[c]{\Checkmark} & \makecell[c]{\XSolidBrush}& \makecell[c]{\XSolidBrush} & \makecell[c]{\Checkmark} & \makecell[c]{\Checkmark} & \makecell[c]{\Checkmark} & \makecell[c]{\Checkmark} & \makecell[c]{\Checkmark} \cr\hline
         \bf Proposed & \makecell[c]{\Checkmark} & \makecell[c]{\Checkmark}& \makecell[c]{\Checkmark} & \makecell[c]{\Checkmark} & \makecell[c]{\Checkmark} & \makecell[c]{\Checkmark} & \makecell[c]{\Checkmark} & \makecell[c]{\Checkmark}\cr\hline
    \Xhline{1.2pt}
    \end{tabular}
\end{table*}

However, it is worth noting that both the aforementioned schemes rely on independent RSS or AOA information to establish fingerprint databases, while overlooking the importance of combining accurate angle and distance information to achieve user positioning. In light of this limitation, the authors in \cite{[10]} proposed a novel joint AOA and RSS-based fingerprint positioning scheme, which combines rich angle and distance information to achieve better positioning performance compared with those positioning schemes considering only RSS or AOA information.
Moreover, due to the poor scalability and generalization ability of the fingerprint databases established in the offline phase, as well as the high computational complexity of fingerprint search in the online phase, these schemes are challenging to implement in practical cell-free mMIMO scenarios.

On a different note, most of existing model-free machine learning-based positioning schemes can significantly improve user positioning accuracy while reducing computational complexity \cite{[8],[15]}. For instance, the authors in \cite{[8]} proposed a machine learning-based fingerprinting positioning scheme consisting of two fully connected neural networks (NNs). Specifically, the first NN is adopted to classify the position coordinates of the UEs, while the second NN is leveraged as a regression module to refine position coordinate estimation, thereby ensuring higher positioning accuracy. However, these machine learning-based fingerprint positioning schemes rely on NN to establish fingerprint databases, which adversely affects scalability and generalization. Meanwhile, their increased reliance on supervised learning poses an emerging challenge in obtaining prior optimal data as training labels, which hinders the implementation of practical scenarios. Hence, effective positioning technology strategies must be developed for cell-free mMIMO systems.

Recently, multi-agent reinforcement learning (MARL) has made significant progress and has achieved numerous successes in solving sequential decision-making problems in various research fields, especially in wireless communication \cite{[32],[33],[34],[40]}. In this context, MARL represents a disruptive method that breaks through the application limitations of conventional positioning methods in terms of scalability, computational complexity, and positioning performance, which holds great promise in addressing the challenges mentioned above \cite{[30]}. In particular, MARL is a generalization of single-agent RL, which enables agents to learn the optimal strategy through interaction with the MARL environment and each other directly. Capitalizing on these benefits, MARL-based solutions have been studied for various wireless resource allocation problems, including dynamic power control \cite{[5]}, antenna selection \cite{[29]}, spectrum access \cite{[31]}, etc. For example, the authors in \cite{[5]} proposed an innovative MARL-based power control scheme, which combines predictive management and distributed optimization architecture to provide a dynamic strategy for solving high-dimensional signal processing problems while reducing computational complexity. Also, the authors in \cite{[31]} designed a collaborative dynamic spectrum access strategy by incorporating federated learning and MARL, which adopts federated learning to promote multiple users to collaborate and optimize a system objective without sharing their training data, thereby improving communication efficiency and enhancing user data privacy. Typically, these MARL-based schemes adopt a single-layer network, which takes all observed states as network inputs to solve optimization problems. However, amalgamating various types of observed states into a unified MARL network may not be conducive to improving system performance. Therefore, it becomes increasingly imperative to design different MARL networks tailored to specific types of observed states to jointly solve optimization problems.
\begin{figure}[t]
\centering
    \includegraphics[scale=0.675]{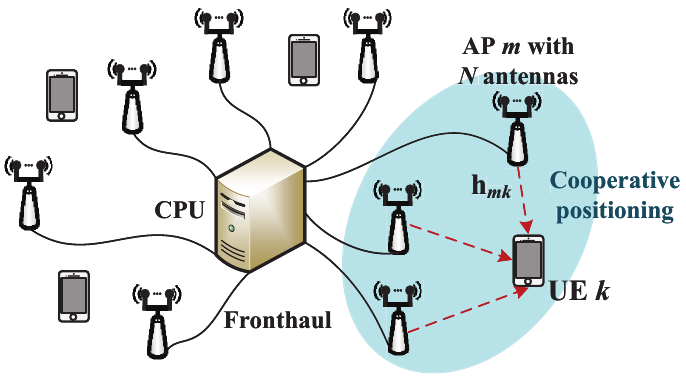}
    \caption{Illustration of a cell-free mMIMO system.
    \label{fig1}}
\end{figure}
\subsection{Motivations and Contributions}
Motivated by the above observations and in contrast to existing approaches (including non-learning-based fingerprint positioning schemes and machine learning-based fingerprint positioning schemes), we initially eliminate the deployment of positioning reference points and the establishment of a fingerprint database in the offline stage to avoid significant computational overhead, poor scalability, and weak generalization ability that limit the actual implementation of the positioning scheme. Then, we introduce a novel MARL network that differentiates from deep learning approaches, aiming to design an effective positioning scheme to address the aforementioned challenges. This innovation reduces the dependence of conventional model-free learning-based positioning schemes on supervised learning, thereby enhancing scalability and real-time performance. Moreover, to the best of the authors' knowledge, there is no research on partial cooperative positioning in mMIMO systems. Among the existing positioning estimation schemes, including three-dimensional convolution NN \cite{[15]} and weighted K-nearest neighbor (WKNN) \cite{[8],[9],[10]}, all APs are selected to participate in user positioning, resulting in partial remote APs affecting their final positioning performance due to evaluation bias. Therefore, we introduce a cooperative WKNN (Co-WKNN)-based positioning estimation scheme to select APs with high correlation to participate in user positioning. The comparisons between our work and several aforementioned existing works are summarized in Table \uppercase\expandafter{\romannumeral1}, and the major contributions of this paper are listed as follows:
\begin{itemize}
\item We first investigate a cell-free mMIMO system and adopt a discrete Fourier transform (DFT) operation to map the original channel information into the angular domain. Then, we derive the angular domain channel power matrix and RSS values for user positioning.

\item We propose a cooperative positioning architecture that combines angle and distance positioning similarity, and analyze the impact of different APs participating in the evaluation on positioning performance to effectively strike a balance between positioning performance and computational complexity.

\item We introduce a joint positioning and correction MARL network to optimize the user positioning problem, which primarily relies on RSS information for initial positioning, supplemented by AOA information for positioning correction, thereby enhancing positioning accuracy. Moreover, we propose a Co-WKNN-based estimation scheme to select APs with high correlation to participate in user positioning for suppressing the impact of biased APs on positioning performance.
\end{itemize}

The rest of this paper is organized as follows. Section \uppercase\expandafter{\romannumeral2} discusses an angular domain channel model for cell-free mMIMO systems, pilot training, channel estimation, and positioning extraction. Section \uppercase\expandafter{\romannumeral3} introduces the joint angle and distance-based cooperative positioning architecture and studies the relationship between the number of participants and positioning performance. Then, in Section \uppercase\expandafter{\romannumeral4}, we propose a novel MARL-based positioning algorithm that combines the preliminary positioning network with a supplementary correction network. In Section \uppercase\expandafter{\romannumeral5}, numerical results, performance analysis, and a comparison of communication overhead for the proposed positioning scheme with conventional fingerprint positioning are provided. Finally, Section \uppercase\expandafter{\romannumeral6} draws the major conclusions and future directions.
\newcounter{mytempeqncnt_1}

\emph{\textbf{{    Notation}}}: The superscripts $\left(\cdot\right)$\textsuperscript{\emph{\textrm{H}}}, $\left(\cdot\right)$\textsuperscript{\emph{$\ast$}} and $\left(\cdot\right)$\textsuperscript{\emph{T}} are utilized to represent the conjugate transpose, conjugate, and transpose, respectively. The matrices and column vectors are denoted by boldface uppercase letters $\bf{X}$ and boldface lowercase letters $\bf{x}$, respectively. $|\cdot|$, $\|\cdot\|$, and $\nabla$ denote the determinant of a matrix, the Euclidean norm, and gradient, respectively. ${\text{tr}\{\cdot\}}$, $\mathbb{E\{\cdot\}}$,  and $\triangleq$ are the trace, expectation, and definitions, respectively. $\odot$ denotes the element-wise products. $\mathbb{R}^n$, and $\mathbb{C}^n$ represent the $n$-dimensional spaces of real and complex numbers, respectively.
Finally, the circularly symmetric complex Gaussian random variable $x$ with variance $\sigma^2$ is denoted by  $x\sim{{\cal N}_\mathbb{C}}\left({0},\sigma^2\right)$.
\section{System Model}
As illustrated in Fig. 1, we consider a cell-free mMIMO system consisting of $K$ single-antenna UEs, $M$ arbitrarily distributed APs equipped with $N$ antennas, and a central processing unit (CPU), and denote $\mathcal{M}=\{1,\ldots,M\}$ and $\mathcal{K}=\{1,\ldots,K\}$. The APs and UEs are arbitrarily distributed in a wide coverage area and each AP is connected via fronthaul connections to a CPU possessing substantial processing capabilities. Moreover, all the APs serve all the UEs utilizing the same time and frequency resources and achieve user positioning through mutual collaboration.
Without loss of generality, the time-division duplex (TDD) protocol having a pilot transmission phase for channel estimation and user positioning is adopted in our system, where the length of each coherence time block is denoted as $\tau_c$ \cite{[1]}. Following the standard TDD protocol \cite{[2]}, each coherence time block $\tau_c$ can be divided into two parts, $\tau_p$ symbols are exploited for uplink pilot transmission phase, and $\tau_u$ symbols are utilized for user positioning phase, satisfying $\tau_c = \tau_p + \tau_u$.
\begin{figure*}[b]
\hrulefill
\normalsize
\setcounter{equation}{0}
\begin{equation}
\begin{split}
\mathbf{h}_{mk}=\sqrt{\frac{\kappa_{mk}}{\kappa_{mk}+1}}\mathbf{h}_{mk}^{\mathrm{LoS}} + \sqrt{\frac{1}{\kappa_{mk}+1}}\mathbf{h}_{mk}^{\mathrm{NLoS}}=\sqrt{\frac{\kappa_{mk}\beta_{mk}}{\kappa_{mk}+1}}\varkappa_{mk}\mathbf{a}(\bar{\theta}_{mk})+\sqrt{\frac{\beta_{mk}}{L_{mk}(\kappa_{mk} + 1)}}\sum_{l=1}^{L_{mk}}\alpha_{mk}^{l}\mathbf{a}(\theta_{mk}^{l}).
\end{split}
\end{equation}
\hrulefill
\normalsize
\setcounter{equation}{4}
\begin{equation}
\begin{split}
\mathbf{y}_{mk}^p \triangleq {\frac{1}{\sqrt{\tau_p}}}\mathbf{Y}_m^p\boldsymbol{\phi}_{k}^\ast = \sum_{i \in \mathcal{K}} {\frac{\sqrt{p_i}}{\sqrt{\tau_p}}}\mathbf{h}_{mi}\boldsymbol{\phi}_{i}^{T}\boldsymbol{\phi}_{k}^\ast + {\frac{1}{\sqrt{\tau_p}}}\mathbf{N}_m\boldsymbol{\phi}_{k}^\ast
=\sum_{i \in \mathcal{P}_k}\sqrt{p_i\tau_p}\mathbf{h}_{mi} + \mathbf{n}_{mk}.
\end{split}
\end{equation}
\label{eq1}
\end{figure*}
\subsection{Channel Model}
In cell-free mMIMO systems, due to the presence of numerous scatterers in the coverage area, wireless signals propagate along multiple paths in practical transmission scenarios. Moreover, we consider a quasi-static block fading model such that within each coherence time block, the channels are frequency flat and static. Therefore, the channel vector $\mathbf{h}_{mk} \in \mathbb{C}^{N \times 1}$ between AP $m$ and UE $k$, with $L_{mk}$ scattering paths \cite{[7]} can be characterized as a Rician fading channel, comprising a semi-deterministic LoS path component and a stochastic NLoS path component, satisfying (1), shown at the bottom of this page, where $\mathbf{h}_{mk}^{\mathrm{LoS}}=\sqrt{\beta{mk}}\varkappa_{mk}\mathbf{a}(\bar{\theta}_{mk}) \in \mathbb{C}^{N \times 1}$ is the deterministic LoS component and $\mathbf{h}_{mk}^{\mathrm{NLoS}}=\sqrt{\frac{\beta{mk}}{L_{mk}}}\sum_{l=1}^{L_{mk}}\alpha_{mk}^{l}\mathbf{a}(\theta_{mk}^{l}) \in \mathbb{C}^{N \times 1}$ is the NLoS component. $\kappa_{mk}$ and $\varkappa_{mk}$ represent the Rician $\kappa$-factor and the phase-shift, respectively.
Moreover, $\beta_{mk}$ and $\alpha_{mk}^{l} \sim {{\mathcal N}_\mathbb{C}}(0,1)$ denote the large-scale fading coefficient and the small-scale fading coefficient of the $l$-th path, $l \in \{1,\dots,L_{mk}\}$, respectively. $\mathbf{a}(\bar{\theta}_{mk}) \in \mathbb{C}^{N \times 1}$ and $\mathbf{a}(\theta_{mk}^{l}) \in \mathbb{C}^{N \times 1}$ represents the array steering vector of LoS and NLoS paths, where the former can be modeled as
\begin{equation}
\setcounter{equation}{2}
\mathbf{a}(\bar{\theta}_{mk}) = \Big[1, e^{-j2\pi\frac{\Delta}{\lambda}\mathrm{cos}(\bar{\theta}_{mk})}, \ldots, e^{-j2\pi\frac{(N-1)\Delta}{\lambda}\mathrm{cos}(\bar{\theta}_{mk})}\Big]^{T},
\label{eq2}
\end{equation}
and the latter can be modeled as
\begin{equation}
\setcounter{equation}{3}
\mathbf{a}(\theta_{mk}^{l}) = \Big[1, e^{-j2\pi\frac{\Delta}{\lambda}\mathrm{cos}(\theta_{mk}^{l})}, \ldots, e^{-j2\pi\frac{(N-1)\Delta}{\lambda}\mathrm{cos}(\theta_{mk}^{l})}\Big]^{T},
\label{eq2}
\end{equation}
where $\Delta$ and $\lambda$ denote the antenna spacing and signal wavelength, respectively. $\bar{\theta}_{mk}$ and $\theta_{mk}^{l} \in [0, \pi]$ are the AOA of LoS and NLoS paths, respectively.
\subsection{Pilot Training and Channel Estimation}
In the uplink pilot training phase, we assume that $\tau_p$ mutually orthogonal pilot sequences $\boldsymbol{\phi}_{1}, \ldots, \boldsymbol{\phi}_{\tau_p}$ with $\boldsymbol{\phi}_{k} \in \mathbb{C}^{\tau_p \times 1}$ of UE $k$ are exploited for channel estimation, satisfying $\|\boldsymbol{\phi}_{k}\|^2=\tau_{p}$. In particular, all the UEs share the same $\tau_p$ mutually orthogonal pilot sequences $\boldsymbol{\phi}_{k}$, $\forall k \in \{1,\ldots,\tau_p\}$. Besides, $\boldsymbol{\phi}_{k}$ is assigned to more than one UE, which promotes a large network with $K > \tau_p$. We denote $\mathcal{P}_k \subset \mathcal{K}$ as the index subset of the UEs which adopt the same pilot sequence $\boldsymbol{\phi}_{k}$ as UE $k$ including itself.

First of all, the received uplink signal $\mathbf{Y}_m^p \in \mathbb{C}^{N \times \tau_p}$ at AP $m$ after all the UEs transmit their pilots can be given as
\begin{equation}
\setcounter{equation}{4}
\mathbf{Y}_m^p = \sum_{i \in \mathcal{K}}\sqrt{p_i}\mathbf{h}_{mi}\boldsymbol{\phi}_{i}^{T} + \mathbf{N}_m,
\label{eq3}
\end{equation}
where $p_i \geqslant 0$ denotes the transmit power of UE $i$ and $\mathbf{N}_m \in \mathbb{C}^{N \times \tau_p}$ is the receiver noise with independent ${{\mathcal N}_\mathbb{C}}(0,\sigma^2)$ entries and the noise power $\sigma^2$.

Then, the AP performs coherent linear processing \cite{[1]} on the received pilot signal $\mathbf{Y}_m^p$. By multiplying the corresponding conjugate pilot sequence $\boldsymbol{\phi}_{k}^\ast$ of UE $k$, the received pilot signal of the UEs in $\mathcal{P}_k$ at AP $m$ can be obtained as (5), shown at the bottom of this page, where $\mathbf{n}_{mk} \triangleq \mathbf{N}_m\boldsymbol{\phi}_{k}^\ast/\sqrt{\tau_p} \sim {{\mathcal N}_\mathbb{C}}(\mathbf{0},\sigma^2\mathbf{I}_N)$ is the resulting noise.

Note that (5) is applicable along with any linear estimator, such as least square (LS) and minimum mean-squared error (MMSE) estimator. Here, we consider the standard LS estimator that minimizes $\|\mathbf{y}_{mk}^p - \sqrt{p_k\tau_p}\mathbf{h}_{mk}\|$, due to its simplicity and low computational complexity \cite{[38]}. Then, the estimate of the channel $\mathbf{h}_{mk}$ can be acquired as
\begin{equation}
\setcounter{equation}{6}
{\mathbf{\hat{h}}}_{mk} = \frac{1}{\sqrt{p_k\tau_p}}\mathbf{y}_{mk}^p=\frac{1}{\sqrt{p_k\tau_p}}\Big(\sum_{i \in \mathcal{P}_k}\sqrt{p_i\tau_p}\mathbf{h}_{mi} + \mathbf{n}_{mk}\Big).
\label{eq5}
\end{equation}
\subsection{Cooperative Positioning Extraction}
During the cooperative positioning extraction phase, to accurately locate the position of all UEs in cell-free mMIMO systems, each AP aims to determine positioning information from the received signal and achieve user positioning through mutual assistance.
In the following, we discuss two commonly adopted approaches for extracting positioning information.
\subsubsection{RSS-based Extraction \cite{[10],[37]}}
Due to the assumption that $K > \tau_p$, the RSS value $\psi_{mk}$ for the multiple links between AP $m$ and UE $k$ is jointly determined by the channel information of all the UEs in the index subset $\mathcal{P}_k$. However, it is not conducive to accurately extracting the positioning information\footnote{In the non-orthogonal pilot scheme, due to the inability of AP $m$ to distinguish simultaneous transmission UEs from set $\mathcal{P}_k$, it is impossible to accurately locate UE $k$. In contrast, in the orthogonal pilot scheme, AP $m$ can extract the RSS value $\psi_{mk}$ of UE $k$ from the received signal, which can be utilized to locate UE $k$.} \cite{[37]}. To improve the accuracy of the extracted positioning information, we assign $K$ mutually orthogonal pilot sequences to all the UEs, i.e., $K = \tau_p$, such that the RSS value $\psi_{mk}$ of UE $k$ is only determined by its channel information $\mathbf{h}_{mk}$, which can be expressed as
\begin{equation}
\setcounter{equation}{7}
\psi_{mk}=\Big\|\sum_{i \in \mathcal{P}_k}\sqrt{p_i\tau_p}\mathbf{h}_{mi}\Big\|^2=p_k\tau_p\|\mathbf{h}_{mk}\|^2.
\label{eq6}
\end{equation}
\begin{figure*}[t]
\centering
    \includegraphics[scale=0.25]{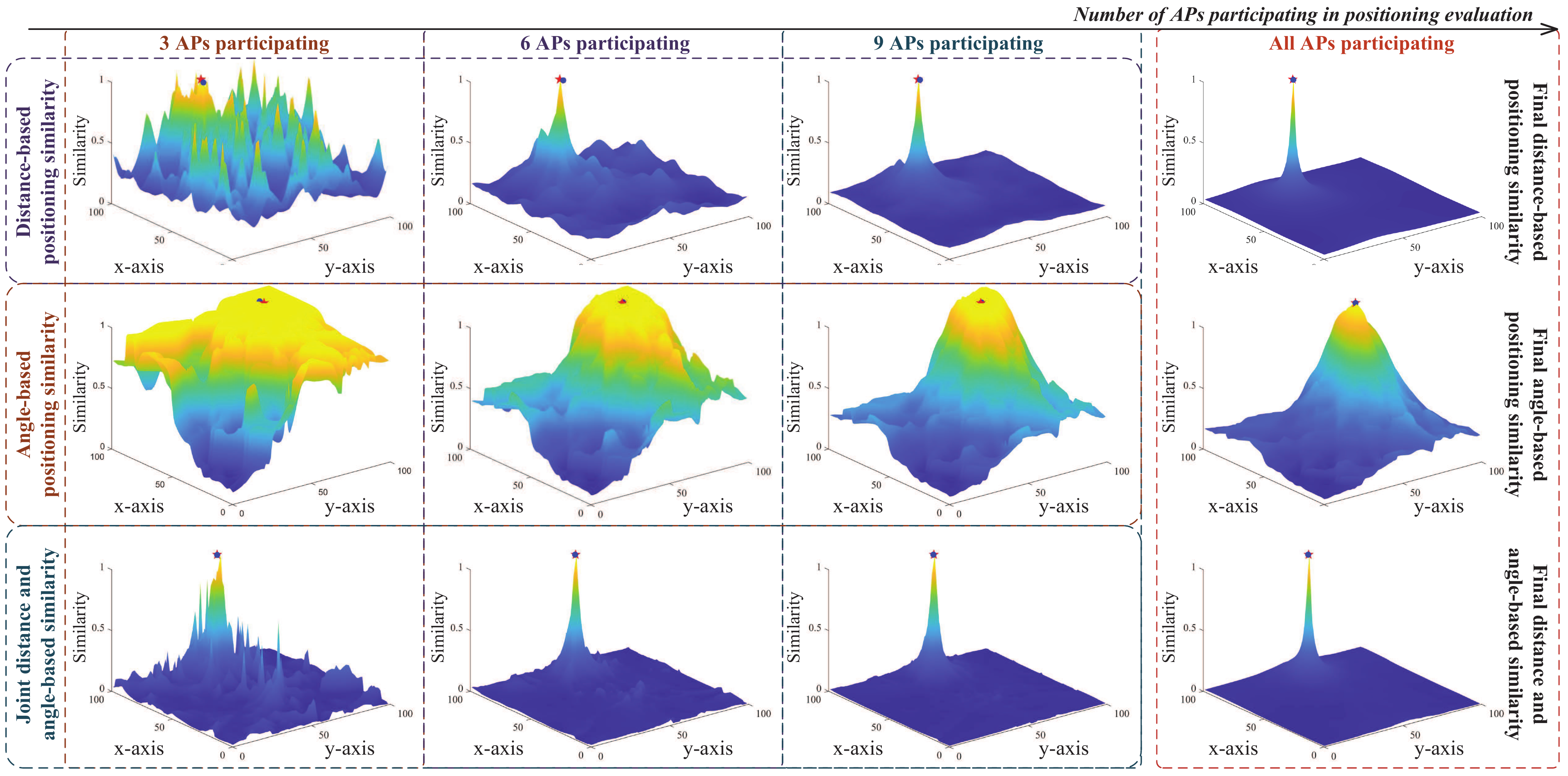}
    \caption{Illustration of positioning similarity coefficient versus the number of APs participating in user positioning, where the red star represents the actual position of the UE, and the blue circle represents the estimated position of the UE.
    \label{fig1}}
\end{figure*}
Moreover, the presence of small-scale fading generally causes random fluctuations in RSS value $\psi_{mk}$.
Therefore, we can apply the channel hardening technique that averages the small-scale fading across multiple slots \cite{[36]} such that the normalized instantaneous channel gain converge to the deterministic average channel gain asymptotically.
Then, based on the assumption above, the RSS value $\psi_{mk}$ can be further simplified as
\begin{equation}
\setcounter{equation}{8}
\psi_{mk} \approx p_k\tau_p \mathbb{E}\Big\{\|\mathbf{h}_{mk}\|^2\Big\} = Np_k\tau_p\beta_{mk},
\label{eq7}
\end{equation}
which shows the RSS value $\psi_{mk}$ is directly proportional to the large-scale fading coefficient $\beta_{mk}$, indicating that it is inversely proportional to the distance $d_{mk}$ between AP $m$ and UE $k$. Correspondingly, the RSS vector related to all the APs with respect to UE $k$ can be expressed as $\mathbf{\Psi}_{k}=[\psi_{1k},\ldots,\psi_{Mk}]^\mathrm{T} \in \mathbb{R}^{M \times 1}$.
\begin{figure*}[b]
\hrulefill
\normalsize
\setcounter{equation}{9}
\begin{equation}
\begin{aligned}
{\Big[\mathbf{G}_k\Big]}_{n,m}&=\sqrt{\frac{\kappa_{mk}\beta_{mk}}{\kappa_{mk}+1}}\varkappa_{mk}e^{-j\frac{N-1}{2}\Big(\frac{2\pi}{N}n+\frac{2\pi\Delta}{\lambda}\mathrm{cos}(\bar{\theta}_{mk})\Big)} \frac{\mathrm{sin}\bigg[\Big(\frac{2\pi}{N}n+\frac{2\pi\Delta}{\lambda}\mathrm{cos}(\bar{\theta}_{mk})\Big)\frac{N}{2}\bigg]}{\mathrm{sin}\bigg[\Big(\frac{2\pi}{N}n+\frac{2\pi\Delta}{\lambda}\mathrm{cos}(\bar{\theta}_{mk})\Big)\frac{1}{2}\bigg]}\\
&+\sqrt{\frac{\beta_{mk}}{L_{mk}(\kappa_{mk}+1)}}\sum_{l=1}^{L_{mk}}\alpha_{mk}^{l}e^{-j\frac{N-1}{2}\Big(\frac{2\pi}{N}n+\frac{2\pi\Delta}{\lambda}\mathrm{cos}(\theta_{mk}^{l})\Big)} \frac{\mathrm{sin}\bigg[\Big(\frac{2\pi}{N}n+\frac{2\pi\Delta}{\lambda}\mathrm{cos}(\theta_{mk}^{l})\Big)\frac{N}{2}\bigg]}{\mathrm{sin}\bigg[\Big(\frac{2\pi}{N}n+\frac{2\pi\Delta}{\lambda}\mathrm{cos}(\theta_{mk}^{l})\Big)\frac{1}{2}\bigg]}.
\end{aligned}
\end{equation}
\end{figure*}
\subsubsection{AOA-based Extraction \cite{[7],[8]}}
Since RSS information contains only rough channel information, it cannot satisfy the highly accurate positioning requirements in complex communication environments. Moreover, compared to conventional RSS-based extraction schemes, AOA information is composed of large-scale statistical channel information, which is related to the spatially sparse structure of the channel \cite{[10]}. Then, we can apply the DFT operation to map the received estimated channel $\mathbf{\hat{h}}_{mk}$ into the angular domain channel $\mathbf{g}_{mk}$. Let the DFT matrix be $\mathbf{F} \in \mathbb{C}^{N \times N}$ and the element in the $i$-th row and $j$-th column of it can be expressed as $[\mathbf{F}]_{i,j}=e^{-j2\pi\frac{(i-1)(j-1)}{N}}$.
Then, by left multiplying a DFT matrix on the estimated channel $\mathbf{\hat{h}}_{mk}$, the angular domain channel response matrix $\mathbf{G}_k \in \mathbb{C}^{N \times M}$ is given by
\begin{equation}
\setcounter{equation}{9}
\mathbf{G}_k=\Big[\mathbf{g}_{1k},\ldots,\mathbf{g}_{Mk}\Big]=\Big[\mathbf{F}\mathbf{\hat{h}}_{1k},\ldots,\mathbf{F}\mathbf{\hat{h}}_{Mk}\Big],
\label{eq8}
\end{equation}
where its element in the $n$-th row and $m$-th column can be given by (10), shown at the bottom of the the page.

In practice, the random fluctuations in time-varying wireless channels caused by small-scale fading pose certain challenges to acquiring accurate channel estimates, making it difficult and computationally expensive to extract the embedded positioning information. To this end, we adopt statistical channel data as a metric since it varies over time slowly in general to suppress the impact of channel fluctuations caused by small-scale fading. Then, the angular domain channel power matrix $\mathbf{\Theta}_k \in \mathbb{R}^{N \times M}$ of UE $k$ can be expressed as \cite{[4]}
\begin{equation}
\setcounter{equation}{11}
\mathbf{\Theta}_k \triangleq \mathbb{E}\Big\{\mathbf{G}_k \odot \mathbf{G}_k^\ast\Big\},
\label{eq10}
\end{equation}
where $[\mathbf{\Theta}_k]_{n,m} \triangleq \mathbb{E}\Big\{\Big|[\mathbf{G}_k]_{n,m}\Big|^2\Big\}$ is the element in the $n$-th row and $m$-th column. It is important to notice that the angular domain channel power matrix $\mathbf{\Theta}_k$ provides an effective description of AOA $\theta_{mk}^{l}$ and channel power distribution in the angular domain, which facilitates the acquisition of precise positioning of cell-free mMIMO systems.
\section{Positioning Model}
In this section, we first investigate the user positioning model and compare various similarity relationships. Then, we analyze the performance of different numbers of APs participating in positioning, as shown in Fig. 2, and propose a cooperative positioning architecture to select the appropriate number of APs.
\begin{figure*}[b]
\hrulefill
\normalsize
\setcounter{equation}{13}
\begin{equation}
\begin{split}
\mathbf{\Xi}^{\mathrm{a}}(\mathbf{\Theta}_k,\hat{\mathbf{\Theta}}_{k})=\frac{1}{\sqrt{M}}\sum_{m \in \mathcal{M}} \mathbf{\Xi}^{\mathrm{a}}_m(\mathbf{\Theta}_k,\hat{\mathbf{\Theta}}_{k})
=\frac{1}{\sqrt{M}}\sum_{m \in \mathcal{M}} \frac{[\mathbf{\Theta}_k]_{:,m}^{T}[\hat{\mathbf{\Theta}}_{k}]_{:,m}}{\|[\mathbf{\Theta}_k]_{:,m}\|\|[\hat{\mathbf{\Theta}}_{k}]_{:,m}\|}.
\end{split}
\end{equation}
\hrulefill
\normalsize
\setcounter{equation}{14}
\begin{equation}
\begin{split}
\mathbf{\bar{\Xi}}^{\mathrm{d}}(\mathbf{\Psi}_k,\hat{\mathbf{\Psi}}_{k})=\frac{\mathbf{\Xi}^{\mathrm{d}}(\mathbf{\Psi}_k,\hat{\mathbf{\Psi}}_{k})}{\max_{\forall i \in \mathcal{K}} \Big\{\mathbf{\Xi}^{\mathrm{d}}(\mathbf{\Psi}_i,\hat{\mathbf{\Psi}}_{i})\Big\}}=\frac{\sqrt{\sum_{m \in \mathcal{M}}|\psi_{mk}-\hat{\psi}_{mk}|^2}}{\max_{\forall i \in \mathcal{K}}\Big\{\sqrt{\sum_{m \in \mathcal{M}}|\psi_{mi}-\hat{\psi}_{mi}|^2}\Big\}}.
\end{split}
\end{equation}
\label{eq1}
\end{figure*}
\begin{figure*}[t]
\centering
    \includegraphics[scale=0.258]{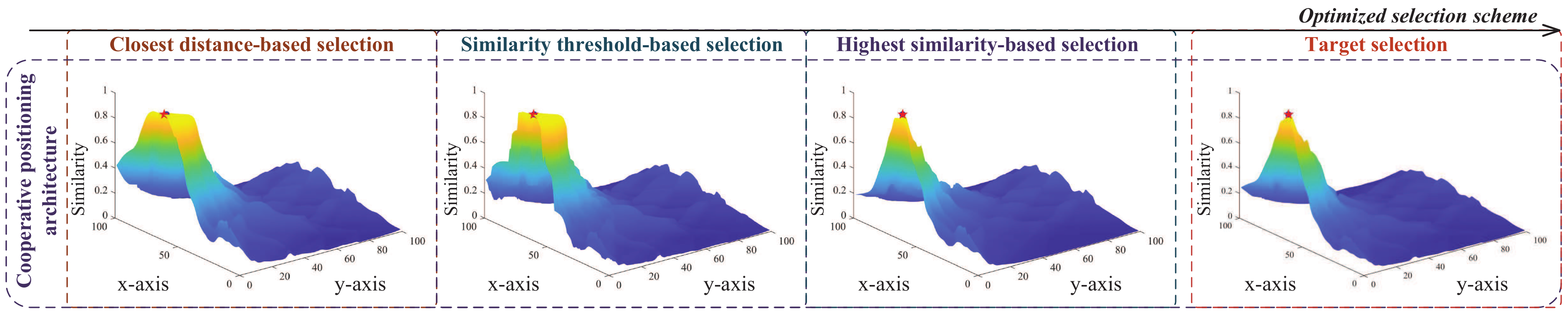}
    \caption{Illustration of positioning similarity coefficient under various estimation schemes.
    \label{fig1}}
\end{figure*}
\subsection{User Positioning Model}
In cell-free mMIMO systems, to quantitatively analyze the positioning accuracy, we denote $(x_k,y_k)$ and $(\hat{x}_{k},\hat{y}_{k})$ as the actual and estimated position, and adopt the root mean square error (RMSE) to evaluate the overall positioning error
\begin{equation}
\setcounter{equation}{12}
e_\text{RMSE}=\sqrt{\frac{1}{K}\sum_{k \in \mathcal{K}}\Big((\hat{x}_{k}-x_k)^2+(\hat{y}_{k}-y_k)^2\Big)}.
\label{eq11}
\end{equation}

It is obvious that to accurately estimate the positions of all UEs, we need to acquire the optimal estimation position within a given region to minimize RMSE $e_\text{RMSE}$. However, the actual position of each UE is unknown in advance, resulting in the inability to achieve user positioning through (12), which only provides a definition of positioning accuracy. Instead, the estimated channel state information of each UE is known in advance.
As an alternative, we adopt similarity criteria to evaluate the correlation between the estimated position and the actual position, thereby achieving user positioning.
\subsection{Positioning Similarity Model}
In this subsection, we adopt the aforementioned RSS or AOA information to define a concept, termed the positioning similarity coefficient to infer the correlation between the actual position $(x_k,y_k)$ and estimated position $(\hat{x}_{k},\hat{y}_{k})$ of UE $k$. Theoretically, the larger the positioning similarity coefficient, the higher the correlation between the actual physical position and the estimated one.

However, due to the impairments of wireless transmission environments, noise, and interference, the similarity coefficients obtained cannot always capture the actual physical position correlation. As illustrated in the first row of Fig. 2, for the distance-based criterion with a small number of participating APs, the similarity coefficient between the actual position and estimated position does not exhibit a monotonically increasing relationship with the Euclidean distance. On the other hand, capturing more accurate AOA information can generally better reflect the actual physical position correlation. However, an excessive number of estimated positions in areas with high similarity can also undermine positioning accuracy.

Therefore, we advocate a combination of the two aforementioned similarity schemes and adopt a joint similarity criterion exploiting an enhanced distance information that utilizes the angle information as a supplement to improve user positioning. Then, the corresponding similarity coefficient can be jointly determined by all APs and given by
\begin{equation}
\setcounter{equation}{13}
\mathbf{\Xi}^{\mathrm{j}}(\mathbf{\Theta}_k,\mathbf{\Psi}_k,\hat{\mathbf{\Theta}}_{k},\hat{\mathbf{\Psi}}_{k})=\frac{1}{\mathbf{\bar{\Xi}}^{\mathrm{d}}(\mathbf{\Psi}_k,\hat{\mathbf{\Psi}}_{k})(1-\mathbf{\Xi}^{\mathrm{a}}(\mathbf{\Theta}_k,\hat{\mathbf{\Theta}}_{k}))},
\label{eq12}
\end{equation}
where $\mathbf{\Xi}^{\mathrm{j}}(\mathbf{\Theta}_k,\mathbf{\Psi}_k,\hat{\mathbf{\Theta}}_{k},\hat{\mathbf{\Psi}}_{k})$ quantifies the joint correlation between angle and distance information. Correspondingly, when the actual position $(x_k,y_k)$ and estimated position $(\hat{x}_{k},\hat{y}_{k})$ of UE $k$ exhibit a high correlation, the joint similarity coefficient tends to be 1, while the opposite tends to be 0. Moreover, $\mathbf{\Xi}^{\mathrm{a}}_m(\mathbf{\Theta}_k,\hat{\mathbf{\Theta}}_{k}) \in [0,1]$ and $\mathbf{\bar{\Xi}}^{\mathrm{d}}(\mathbf{\Psi}_k,\hat{\mathbf{\Psi}}_{k}) \in [0,1]$ are the angle-based and the normalized distance-based similarity coefficient, e.g., (14) and (15), shown at the bottom of this page, where $\mathbf{\Xi}^{\mathrm{d}}(\mathbf{\Psi}_k,\hat{\mathbf{\Psi}}_{k})$ and ${\max_{\forall i \in \mathcal{K}} \{\mathbf{\Xi}^{\mathrm{d}}(\mathbf{\Psi}_i,\hat{\mathbf{\Psi}}_{i})\}}$ are the distance-based similarity coefficient between the actual and estimated position and the maximum similarity coefficient, respectively.
\begin{figure*}[b]
\hrulefill
\normalsize
\setcounter{equation}{16}
\begin{equation}
\begin{split}
\Big\{\mathbf{I}_{\mathrm{sort}}\Big(\mathbf{\Xi}_:^{\mathrm{j}}(\mathbf{\Theta}_k,\mathbf{\Psi}_k,\hat{\mathbf{\Theta}}_{k},\hat{\mathbf{\Psi}}_{k})_{1}\Big), \ldots,\mathbf{I}_{\mathrm{sort}}\Big(\mathbf{\Xi}_:^{\mathrm{j}}(\mathbf{\Theta}_k,\mathbf{\Psi}_k,\hat{\mathbf{\Theta}}_{k},\hat{\mathbf{\Psi}}_{k})_{L_{k}}\Big)\Big\}.
\end{split}
\end{equation}
\hrulefill
\normalsize
\setcounter{equation}{20}
\begin{equation}
\begin{split}
\mathbf{\bar{\Xi}}_{\mathcal{S}_{k}^{\mathrm{c}}}^{\mathrm{d}}(\mathbf{\Psi}_k,\hat{\mathbf{\Psi}}_{k})=\frac{\mathbf{{\Xi}}_{\mathcal{S}_{k}^{\mathrm{c}}}^{\mathrm{d}}(\mathbf{\Psi}_k,\hat{\mathbf{\Psi}}_{k})}{\max_{\forall i \in \mathcal{K}}\Big\{\mathbf{{\Xi}}_{\mathcal{S}_{k}^{\mathrm{c}}}^{\mathrm{d}}(\mathbf{\Psi}_i,\hat{\mathbf{\Psi}}_{i})\Big\}}=\frac{\sqrt{\sum_{m \in {\mathcal{S}_{k}^{\mathrm{c}}}}|\psi_{mk}-\hat{\psi}_{mk}|^2}}{\max_{\forall i \in \mathcal{K}}\Big\{\sqrt{\sum_{m \in {\mathcal{S}_{k}^{\mathrm{c}}}}|\psi_{mi}-\hat{\psi}_{mi}|^2}\Big\}}.
\end{split}
\end{equation}
\label{eq1}
\end{figure*}

Note that (14) can capture the correlation between the AOA information extracted from two different positions $(x_k,y_k)$ and $(\hat{x}_{k},\hat{y}_{k})$ of UE $k$, where $\mathbf{\Xi}^{\mathrm{a}}_m(\mathbf{\Theta}_k,\hat{\mathbf{\Theta}}_{k}) \rightarrow 1$ indicates a high correlation between the two, and $\mathbf{\Xi}^{\mathrm{a}}_m(\mathbf{\Theta}_k,\hat{\mathbf{\Theta}}_{k}) \rightarrow 0$, otherwise. Similarly, (15) reflects that the distance-based similarity coefficient $\mathbf{\bar{\Xi}}^{\mathrm{d}}(\mathbf{\Psi}_k,\hat{\mathbf{\Psi}}_{k})$ decreases with an increasing of the correlation between the RSS information, which is the opposite of the angle-based similarity coefficient $\mathbf{\Xi}^{\mathrm{a}}(\mathbf{\Theta}_k,\hat{\mathbf{\Theta}}_{k})$. To unify the similarity criteria, we adopt the reciprocal of the distance-based similarity coefficient as the similarity criterion, where $1/{\mathbf{\bar{\Xi}}^{\mathrm{d}}(\mathbf{\Psi}_k,\hat{\mathbf{\Psi}}_{k})} \rightarrow 1$ indicates a high correlation between the two, and $1/{\mathbf{\bar{\Xi}}^{\mathrm{d}}(\mathbf{\Psi}_k,\hat{\mathbf{\Psi}}_{k})} \rightarrow 0$, otherwise.

Therefore, to achieve user positioning for $m \in \mathcal{M}$, we can transform the original minimizing RMSE into maximizing joint positioning similarity problem, which can be modeled as
\begin{equation}
\setcounter{equation}{16}
\begin{aligned}
\max_{\{\hat{\mathbf{\Theta}}_{k},\hat{\mathbf{\Psi}}_{k}:\forall k\}} \quad &\sum_{k \in \mathcal{K}} \mathbf{\Xi}^{\mathrm{j}}(\mathbf{\Theta}_k,\mathbf{\Psi}_k,\hat{\mathbf{\Theta}}_{k},\hat{\mathbf{\Psi}}_{k}),\\
\quad \mbox{s.t.} \qquad \,\, & \, \hat{\mathbf{\Theta}}_{k} \succ \mathbf{0},\hat{\mathbf{\Psi}}_{k} \succ \mathbf{0}, k \in \mathcal{K},
\label{eq15}
\end{aligned}
\end{equation}
where $\mathbf{\Theta}_k$ and $\mathbf{\Psi}_k$ of UE $k$ are known, we can achieve user positioning by finding the optimal $\hat{\mathbf{\Theta}}_{k}$ and $\hat{\mathbf{\Psi}}_{k}$.
\subsection{Cooperative Positioning Architecture}
Although conventional joint positioning schemes consisting of all APs can provide high positioning performance, their high computational complexity seriously limits the practical feasibility of cell-free mMIMO systems. On the other hand, although relying solely on a single AP for evaluating the positioning similarity coefficient can substantially reduce computational complexity, the limitations of its observed information seriously jeopardize positioning accuracy.
As shown in Fig. 2, we can observe that as the number of APs participating in user positioning increases, the obtained positioning similarity gradually approaches that achieved by the conventional joint positioning schemes consisting of all APs. Meanwhile, through the cooperation among APs, the similarity results tend to converge towards a single tip, matching the actual position of the UE to assist in achieving high-precision user positioning.

Therefore, it is crucial to design a cooperative positioning architecture that considers positioning similarity adaptively select subset of APs, aiming to strike an excellent balance between positioning accuracy and communication overhead. For the AP selection adopted to achieve user positioning, we consider a widely accepted overlapped cooperative mode, where each AP can perform positioning similarity evaluation for multiple different UEs and compare several typical selection schemes, e.g., the closest distance, similarity threshold, and highest similarity \cite{[10]}, as shown in Fig. 3. Compared to the closest distance-based and similarity threshold-based selection schemes, we can observe that the highest similarity-based selection scheme fully exploits the angle and distance information offered by AOA-based and RSS-based similarity and eliminates the AP with weaker correlation to improve positioning accuracy.
Here, we consider the highest similarity-based selection scheme due to its superior performance and denote $\mathcal{S}_{k}^{\mathrm{c}} \subset \mathcal{M}$ as the evaluation subset of UE $k$, which adopts suitable APs to evaluate positioning similarity coefficient, and $L_{k} = |\mathcal{S}_{k}^{\mathrm{c}}|$ as the number of APs in the evaluation subset. Then, the $L_{k}$ APs with the highest similarly that are most relevant to the estimated physical position $(\hat{x}_{k},\hat{y}_{k})$ of UE $k$ can be utilized as its evaluation subset, e.g., (17), shown at the bottom of this page, where $\mathbf{I}_{\mathrm{sort}}(\mathbf{\Xi}_m^{\mathrm{j}}(\mathbf{\Theta}_k,\mathbf{\Psi}_k,\hat{\mathbf{\Theta}}_{k},\hat{\mathbf{\Psi}}_{k}))$ is the position of the coefficient $\mathbf{\Xi}_m^{\mathrm{j}}(\mathbf{\Theta}_k,\mathbf{\Psi}_k,\hat{\mathbf{\Theta}}_{k},\hat{\mathbf{\Psi}}_{k})$ at AP $m$ in the descending sort of the set $\{\mathbf{\Xi}_m^{\mathrm{j}}(\mathbf{\Theta}_k,\mathbf{\Psi}_k,\hat{\mathbf{\Theta}}_{k},\hat{\mathbf{\Psi}}_{k}),m \in \mathcal{M}\}$, where the $m$-th joint similarity coefficient can be expressed as
\begin{equation}
\setcounter{equation}{18}
\mathbf{\Xi}_m^{\mathrm{j}}(\mathbf{\Theta}_k,\mathbf{\Psi}_k,\hat{\mathbf{\Theta}}_{k},\hat{\mathbf{\Psi}}_{k})=\frac{1}{\mathbf{\bar{\Xi}}_m^{\mathrm{d}}(\mathbf{\Psi}_k,\hat{\mathbf{\Psi}}_{k})(1-\mathbf{\Xi}_m^{\mathrm{a}}(\mathbf{\Theta}_k,\hat{\mathbf{\Theta}}_{k}))}.
\label{eq17}
\end{equation}

Therefore, the cooperative positioning coefficient under the highest similarity-based selection $\mathcal{S}_{k}^{\mathrm{c}}$ can be expressed as
\begin{equation}
\setcounter{equation}{19}
\mathbf{\Xi}_{\mathcal{S}_{k}^{\mathrm{c}}}^{\mathrm{j}}(\mathbf{\Theta}_k,\mathbf{\Psi}_k,\hat{\mathbf{\Theta}}_{k},\hat{\mathbf{\Psi}}_{k})=
\frac{1}{\mathbf{\bar{\Xi}}_{\mathcal{S}_{k}^{\mathrm{c}}}^{\mathrm{d}}(\mathbf{\Psi}_k,\hat{\mathbf{\Psi}}_{k})(1-\mathbf{\Xi}_{\mathcal{S}_{k}^{\mathrm{c}}}^{\mathrm{a}}(\mathbf{\Theta}_k,\hat{\mathbf{\Theta}}_{k}))},
\label{eq18}
\end{equation}
where $\mathbf{\Xi}_{\mathcal{S}_{k}^{\mathrm{c}}}^{\mathrm{j}}(\mathbf{\Theta}_k,\mathbf{\Psi}_k,\hat{\mathbf{\Theta}}_{k},\hat{\mathbf{\Psi}}_{k})$ quantifies the joint correlation between angle and distance information under the cooperative positioning architecture. Similarly, when the actual and estimated position have a high correlation, the joint similarity coefficient tends to be 1, while the opposite tends to be 0.
Moreover, the angle-based similarity coefficient adopting the cooperative positioning architecture can be rewritten as
\begin{equation}
\setcounter{equation}{20}
\mathbf{\Xi}_{\mathcal{S}_{k}^{\mathrm{c}}}^{\mathrm{a}}(\mathbf{\Theta}_k,\hat{\mathbf{\Theta}}_{k})=\frac{1}{\sqrt{|\mathcal{S}_{k}^{\mathrm{c}}|}}\sum_{m \in \mathcal{S}_{k}^{\mathrm{c}}}\frac{[\mathbf{\Theta}_k]_{:,m}^{T}[\hat{\mathbf{\Theta}}_{k}]_{:,m}}{\|[\mathbf{\Theta}_k]_{:,m}\|\|[\hat{\mathbf{\Theta}}_{k}]_{:,m}\|},
\label{eq19}
\end{equation}
and the normalized distance-based similarity coefficient can be rewritten as (21), shown at the bottom of this page.
Correspondingly, for $m \in {\mathcal{S}_{k}^{\mathrm{c}}}$, the positioning optimization problem in (16) can be rewritten as
\begin{equation}
\setcounter{equation}{22}
\begin{aligned}
\max_{\{\hat{\mathbf{\Theta}}_{k},\hat{\mathbf{\Psi}}_{k}:\forall k\}} \quad &\sum_{k\in\mathcal{K}}\mathbf{\Xi}_{\mathcal{S}_{k}^{\mathrm{c}}}^{\mathrm{j}}(\mathbf{\Theta}_k,\mathbf{\Psi}_k,\hat{\mathbf{\Theta}}_{k},\hat{\mathbf{\Psi}}_{k}),\\
 \mbox{s.t.} \qquad \,\, & \,\,\, \hat{\mathbf{\Theta}}_{k} \succ \mathbf{0},\hat{\mathbf{\Psi}}_{k} \succ \mathbf{0}, k \in \mathcal{K},\\
& \quad \, 0 < L_{k} \leqslant M, k \in \mathcal{K}.
\label{eq21}
\end{aligned}
\end{equation}

It is obvious that the optimization problem in (22) is non-convex and the computational complexity of conventional fingerprint positioning methods \cite{[7],[8],[9],[10],[11],[16]} is prohibitively high, rendering the original solutions incompatible with cell-free mMIMO systems. Therefore, in the following section, we introduce a novel MARL-based method that overcomes the aforementioned challenges.
\begin{figure*}[t]
\centering
    \includegraphics[scale=0.71]{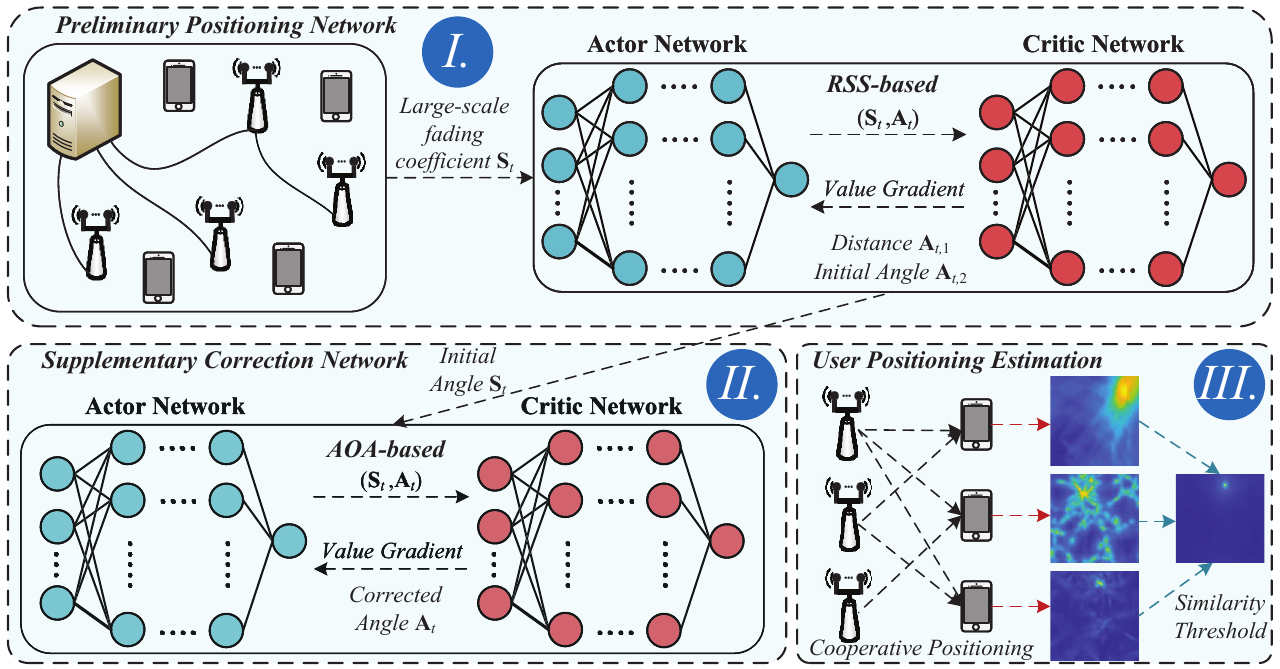}
    \caption{The framework of the proposed MARL-based positioning system.
    \label{fig1}}
\end{figure*}
\section{Proposed MARL-based Positioning Scheme}
In this section, we propose a novel positioning scheme using a MARL network for cell-free mMIMO systems that primarily relies on RSS information for initial positioning, augmented by AOA information for refinement, termed joint positioning and correction (JPC)-multi-agent deep deterministic policy gradient (MADDPG) algorithm.
\subsection{Markov Decision Process Model}
Recently, MARL has attracted much attention for acquiring effective solutions to resource allocation problems in cell-free mMIMO systems \cite{[5],[31]}. Most conventional schemes focus on supervised learning, which requires dense training data incurring high computational complexity. By contrast, MARL is a disruptive method that bypasses the need for prior training data, making it suitable for dynamic wireless environments. In particular, numerous efficient algorithms have been derived, such as MADDPG, where all goal-oriented agents obtain feedback through continuous interaction with the environment to formulate effective strategies.

\begin{figure*}[b]
\hrulefill
\normalsize
\setcounter{equation}{22}
\begin{equation}
\begin{split}
r_{t,m}^{\mathrm{p}}=\sum_{{k} \in \mathcal{S}_{m}^{\mathrm{c,p}}} \mathbf{\bar{\Xi}}_{\mathcal{S}_{{k}}^{\mathrm{c,p}}}^{\mathrm{d}}(\mathbf{\Psi}_k,\hat{\mathbf{\Psi}}_{k}^{m,\mathrm{p}})=\sum_{{k} \in \mathcal{S}_{m}^{\mathrm{c,p}}}\frac{\sqrt{\sum_{m \in {\mathcal{S}_{{k}}^{\mathrm{c,p}}}}|\psi_{mk}-\hat{\psi}_{mk}^{m,\mathrm{p}}|^2}}{\max_{\forall i \in \mathcal{S}_{m}^{\mathrm{c,p}}}\Big\{\sqrt{\sum_{m \in {\mathcal{S}_{{k}}^{\mathrm{c,p}}}}|\psi_{mi}-\hat{\psi}_{mi}^{m,\mathrm{p}}|^2}\Big\}}.
\end{split}
\end{equation}
\hrulefill
\normalsize
\setcounter{equation}{26}
\begin{equation}
\begin{split}
r_{t,m}^{\mathrm{c}}=\sum_{{k} \in \mathcal{S}_{m}^{\mathrm{c,c}}} \mathbf{{\Xi}}_{\mathcal{S}_{{k}}^{\mathrm{c,c}}}^{\mathrm{j}}(\mathbf{\Psi}_k,\mathbf{\Theta}_k,\hat{\mathbf{\Psi}}_{k}^{m,\mathrm{c}},\hat{\mathbf{\Theta}}_{k}^{m,\mathrm{c}})=\sum_{{k} \in \mathcal{S}_{m}^{\mathrm{c,c}}}\frac{1}{\mathbf{\bar{\Xi}}_{\mathcal{S}_{k}^{\mathrm{c,c}}}^{\mathrm{d}}(\mathbf{\Psi}_k,\hat{\mathbf{\Psi}}_{k}^{m,\mathrm{c}})\Big(1-\mathbf{\Xi}_{\mathcal{S}_{k}^{\mathrm{c,c}}}^{\mathrm{a}}(\mathbf{\Theta}_k,\hat{\mathbf{\Theta}}_{k}^{m,\mathrm{c}})\Big)}.
\end{split}
\end{equation}
\label{eq1}
\end{figure*}
In general, MARL is studied by means of Markov decision process (MDP) characterized by a tuple $<\mathcal{S}, \mathcal{A}, \mathcal{P}, \mathcal{R},  \gamma>$, where $\mathcal{S}$ and $\mathcal{A}$ are the observed state space and the assigned action space of each agent, respectively. Variables $\mathcal{P}:(\mathcal{S},\mathcal{A})\rightarrow\mathcal{S}$ and $\mathcal{R}$ represent the state transition function and the expected
reward, respectively. Besides, $\gamma$ denotes the discounted factor of the reward. Moreover, MARL algorithms can be divided into three implementation forms: value-based, policy-based, and actor-critic methods. Compared with value-based and policy-based methods, the critic network in actor-critic methods fits the action-value function in a continuous space, and the action network does not require optimal strategy search in discrete space, indicating that this architecture is particularly appealing for continuous optimization problems, such as MADDPG. Therefore, we conceptualize the studied cell-free mMIMO system as a multi-agent system and adopt the MADDPG algorithm, composed of a policy-based actor network and a value-based critic network, as the underlying architecture to achieve user positioning. Meanwhile, the adoption of the MADDPG algorithm also belongs to centralized training and decentralized execution (CTDE) mechanisms, where the policy-based actor network completes decentralized execution, while the value-based critic network completes centralized training. Specifically, in the centralized phase, we gather data from all agents, enabling us to optimize policies and learning processes from a global perspective. In contrast, in the decentralized phase, each agent makes independent decisions based on its strategy, without relying on the information or control of other agents, which can reduce the communication and computational burden during online execution. Although each agent operates independently, due to the consideration of global optimization during the training phase, the strategies of each agent are typically well-coordinated, thereby jointly driving the system towards excellent global performance.
\subsection{Joint Positioning and Correction Network}
In conventional fingerprint positioning methods, a fingerprint database is established by extracting RSS or AOA feature information of reference points to create fingerprints during the offline stage. The accuracy of positioning is closely related to the number of collected reference points. However, these conventional fingerprint positioning methods face challenges such as poor scalability, high computational complexity, and increased communication overhead. These issues make them difficult to achieve in large-scale multi-target positioning scenarios. Therefore, the trend is to develop a novel positioning architecture with higher real-time performance, better scalability, and lower computational complexity.

In this subsection, aligning with the development trend of the positioning scheme mentioned above, we introduce a novel joint positioning and correction network with MADDPG to achieve user positioning, namely JPC-MADDPG, which consists of the preliminary positioning network and the supplementary correction network, as shown in Fig. 4.
\begin{figure*}[t]
\centering
    \includegraphics[scale=0.25]{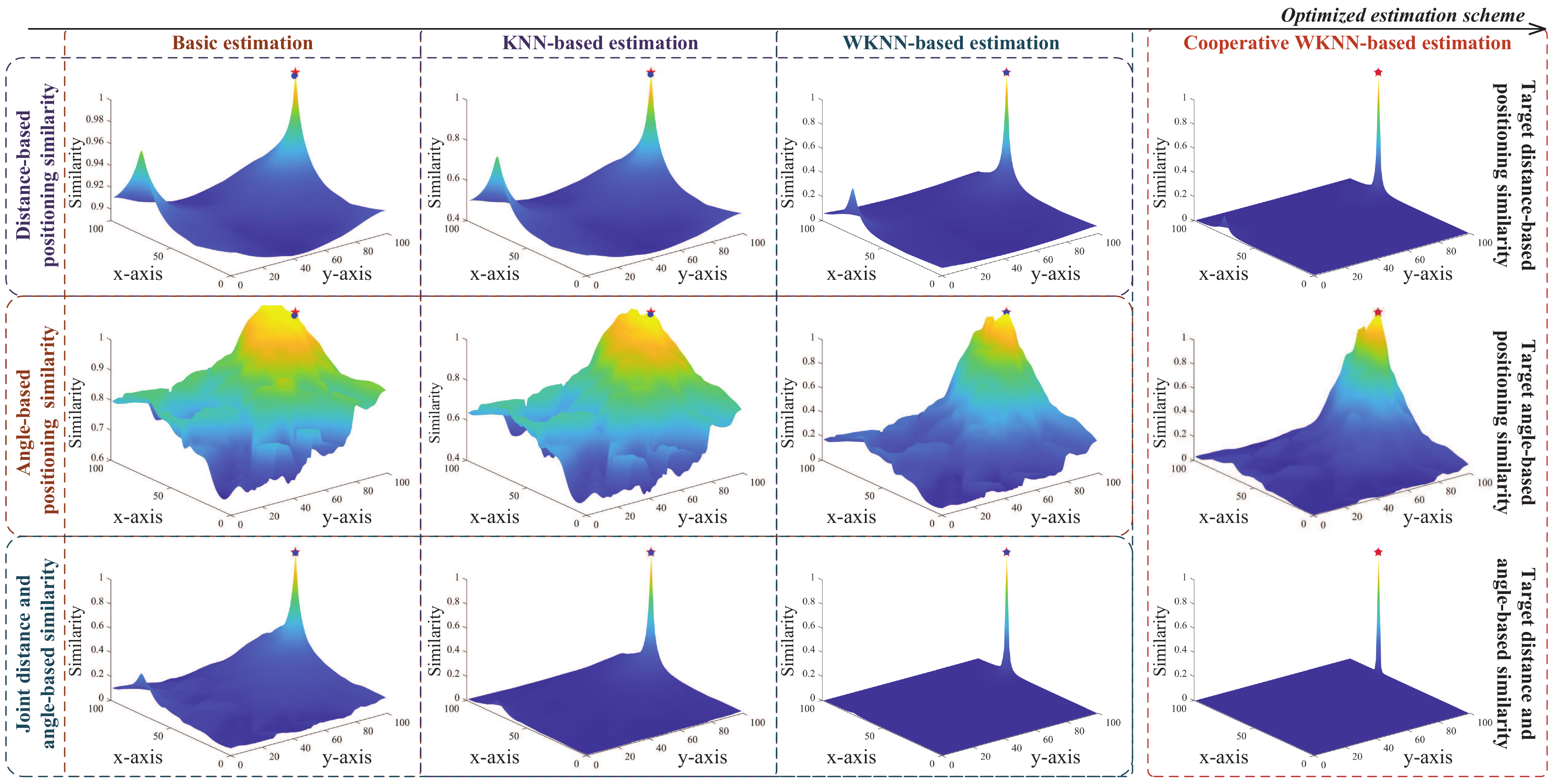}
    \caption{Illustration of positioning similarity coefficient under various estimation schemes, where the basic estimation represents directly selecting the point with the highest similarity coefficient as the estimated position.}
    \label{fig1}
\end{figure*}
\subsubsection{RSS-based Preliminary Positioning Network}
We define all $M$ AP in cell-free mMIMO systems as agents, while all antennas deployed by the same AP are considered as a whole for analysis. In this positioning network, each agent maps its observed RSS information $\mathbf{S}_{t}^{\mathrm{p}}=[\boldsymbol{s}_{t,1}^{\mathrm{p}},\ldots,\boldsymbol{s}_{t,M}^{\mathrm{p}}] \in \mathbb{C}^{K \times M}$ in conjunction with $\boldsymbol{s}_{t,m}^{\mathrm{p}} = [{\psi}_{m1},\ldots,{\psi}_{mK}]^{T} \in \mathbb{C}^{K \times 1}$ to corresponding estimated distance and angle information $\mathbf{A}_{t}^{\mathrm{p}}=[\boldsymbol{a}_{t,1}^{\mathrm{p}},\ldots,\boldsymbol{a}_{t,M}^{\mathrm{p}}] \in \mathbb{C}^{2K \times M}$ in conjunction with $\boldsymbol{a}_{t,m}^{\mathrm{p}} = [\hat{d}_{m1},\ldots,\hat{d}_{mK},\hat{\theta}_{m1}^{\mathrm{p}},\ldots,\hat{\theta}_{mK}^{\mathrm{p}}]^{T} \in \mathbb{C}^{2K \times 1}$ at $t$ time slot, e.g., $\boldsymbol{a}_{t,m}^{\mathrm{p}}=\pi_{k}^{\mathrm{p}}(\boldsymbol{s}_{t,m}^{\mathrm{p}}|m \in \mathcal{M})$.
Considering that this preliminary positioning network locates each UE based on the observed RSS by each AP, we can adopt the updated normalized distance-based positioning similarity coefficient $\mathbf{\bar{\Xi}}_{\mathcal{S}_{k}^{\mathrm{c}}}^{\mathrm{d}}(\mathbf{\Psi}_k,\hat{\mathbf{\Psi}}_{k})$ under the cooperative architecture to quantify the reward of this positioning network $\boldsymbol{r}_{t}^{\mathrm{p}} = [r_{t,1}^{\mathrm{p}},\ldots,r_{t,M}^{\mathrm{p}}]$ with $r_{t,m}^{\mathrm{p}}$ at AP $m$, which can be modeled as (23), shown at the bottom of this page,
where $\mathcal{S}_{m}^{\mathrm{c,p}}$ denotes the evaluation subset of user positioning at AP $m$, which belongs to the mapping matrix of ${\mathcal{S}_{{k}}^{\mathrm{c,p}}}$ under the RSS-based preliminary positioning network. $\hat{\mathbf{\Psi}}_{k}^{m,\mathrm{p}}$ represents the RSS value extracted from the position of UE $k$ estimated by AP $m$, which is determined by the initial position of AP $m$ and the actions output by the positioning network, i.e. $x_{km}^{\mathrm{u,p}} = x_{m}^{\mathrm{a}} + \hat{d}_{mk}\mathrm{cos}(\hat{\theta}_{mk}^{\mathrm{p}})$ and $y_{km}^{\mathrm{u,p}} = y_{m}^{\mathrm{a}} + \hat{d}_{mk}\mathrm{sin}(\hat{\theta}_{mk}^{\mathrm{p}})$.

Then, the policy gradient of the RSS-based preliminary positioning network for $\pi_{m}^{\mathrm{p}}$ can be modeled as
\begin{equation}
\setcounter{equation}{24}
\begin{aligned}
\nabla_{\theta_{\pi_m^\mathrm{p}}}J(\theta_{\pi_m^{\mathrm{p}}})=
{\mathbb{E}
\Big[\nabla_{\theta_{\pi_m^\mathrm{p}}}\pi_m^\mathrm{p}(\boldsymbol{a}_{t,m}^{\mathrm{p}})
Q_{\theta_{Q_{\pi_m^\mathrm{p}}}}(\mathbf{S}_{t}^{\mathrm{p}},\mathbf{A}_{t}^{\mathrm{p}})\Big]},
\label{eq23}
\end{aligned}
\end{equation}
where $\mathcal{D}^{\mathrm{p}}$ denote the experience extraction replay buffer of the preliminary positioning network that stores experience $<\mathbf{S}_{t}^{\mathrm{p}}, \mathbf{A}_{t}^{\mathrm{p}}, \boldsymbol{r}_{t}^{\mathrm{p}}, \mathbf{S}_{t+1}^{\mathrm{p}}>$.

Besides, considering that the action value $Q_{\theta_{Q_{\pi_m^\mathrm{p}}}}(\mathbf{S}_{t}^{\mathrm{p}},\mathbf{A}_{t}^{\mathrm{p}})$ is calculated by the current critic network ${{\theta_{Q_{\pi_m^\mathrm{p}}}}}$, then the mean-squared Bellman error function \cite{[5]} of the current critic network $L({{\theta_{Q_{\pi_m^\mathrm{p}}}}})$ can be defined as
\begin{equation}
\setcounter{equation}{25}
\begin{split}
L({{\theta_{Q_{\pi_m^\mathrm{p}}}}}) = \mathbb{E}_{\mathbf{S}_{t}^{\mathrm{p}},\mathbf{A}_{t}^{\mathrm{p}}\sim \mathcal{D}^{\mathrm{p}}}\bigg[\Big(Q_{{\theta_{Q_{\pi_m^\mathrm{p}}}}}(\mathbf{S}_{t}^{\mathrm{p}},\mathbf{A}_{t}^{\mathrm{p}})-y_{t,m}^{\mathrm{p}}\Big)^2\bigg],
\label{eq24}
\end{split}
\end{equation}
where $y_{t,m}^{\mathrm{p}}=r_{t,m}^{\mathrm{p}}+\gamma^{\mathrm{p}}\Big(Q_{{\theta_{Q_{\pi_m^\mathrm{p,tar}}}}}(\mathbf{S}_{t}^{\mathrm{p,tar}},\mathbf{A}_{t}^{\mathrm{p,tar}})\Big)$ is the target value of agent $k$ with the target critic network value $Q_{{\theta_{Q_{\pi_m^\mathrm{p,tar}}}}}(\mathbf{S}_{t}^{\mathrm{p,tar}},\mathbf{A}_{t}^{\mathrm{p,tar}})$.

Furthermore, to ensure that the target actor network ${\theta_{\pi_m^\mathrm{p,tar}}}$ and critic network ${{\theta_{Q_{\pi_m^\mathrm{p,tar}}}}}$ remains stable, the soft update is carried out with $\tau^\mathrm{p} \ll 1$, and the target network of the preliminary positioning network can be given by
\begin{equation}
\setcounter{equation}{26}
\begin{split}
\left \{
\begin{array}{ll}
{\theta_{\pi_m^\mathrm{p,tar}}} \leftarrow \tau^\mathrm{p}{\theta_{\pi_m^\mathrm{p,tar}}}+(1-\tau^\mathrm{p}){\theta_{\pi_m^\mathrm{p}}},\\
{{\theta_{Q_{\pi_m^\mathrm{p,tar}}}}} \leftarrow \tau^\mathrm{p}{{\theta_{Q_{\pi_m^\mathrm{p,tar}}}}} + (1-\tau^\mathrm{p}){{\theta_{Q_{\pi_m^\mathrm{p}}}}}.
\end{array}
\right.
\label{eq25}
\end{split}
\end{equation}
\begin{figure*}[b]
\hrulefill
\normalsize
\setcounter{equation}{31}
\begin{equation}
\begin{split}
\mathcal{M}_k^{\mathrm{CoW}} =
\Big\{\mathbf{I}_{\mathrm{sort}}\Big({\mathbf{{\Xi}}_{\mathcal{S}_{{k}}^{\mathrm{c}},:}^{\mathrm{j}} (\mathbf{\Psi}_k,\mathbf{\Theta}_k,\hat{\mathbf{\Psi}}_{k}^m,\hat{\mathbf{\Theta}}_{k}^m)_{1}}\Big), \ldots, \mathbf{I}_{\mathrm{sort}}\Big({\mathbf{{\Xi}}_{\mathcal{S}_{{k}}^{\mathrm{c}},:}^{\mathrm{j}} (\mathbf{\Psi}_k,\mathbf{\Theta}_k,\hat{\mathbf{\Psi}}_{k}^m,\hat{\mathbf{\Theta}}_{k}^m)_{M_\mathrm{c}}}\Big)\Big\}_{\mathbf{\Xi}_{\mathcal{S}_{{k}}^{\mathrm{c}},:}^{\mathrm{j}} \geqslant \mathbf{\Xi}_\mathrm{thres}}.
\end{split}
\end{equation}
\label{eq1}
\end{figure*}
\subsubsection{AOA-based Supplementary Correction Network}
Due to the fact that the RSS-based preliminary positioning network relies on extracted RSS information to locate all the UEs, which is proportional to the large-scale fading coefficient, it can only reflect accurate distance information and cannot extract accurate angle information. Therefore, in contrast to the RSS-based preliminary positioning network, this supplementary correction network mainly relies on AOA information to appropriately correct the angle information output of the positioning network. Moreover, this correction network still considers all APs as independent agents, with the difference being that it maps the angle information output by the positioning network to the corrected angle information and optimizes the correction network by combining AOA and RSS information. Then, we can define the observed state and assigned action of the correction network at $t$ time slot as $\mathbf{S}_{t}^{\mathrm{c}}=[\boldsymbol{s}_{t,1}^{\mathrm{c}},\ldots,\boldsymbol{s}_{t,M}^{\mathrm{c}}] \in \mathbb{C}^{K \times M}$ with $\boldsymbol{s}_{t,m}^{\mathrm{c}} = [{\hat{\theta}}_{m1}^{\mathrm{p}},\ldots,{\hat{\theta}}_{mK}^{\mathrm{p}}]^{T} \in \mathbb{C}^{K \times 1}$ and $\mathbf{A}_{t}^{\mathrm{c}}=[\boldsymbol{a}_{t,1}^{\mathrm{c}},\ldots,\boldsymbol{a}_{t,M}^{\mathrm{c}}] \in \mathbb{C}^{K \times M}$ with $\boldsymbol{a}_{t,m}^{\mathrm{c}} = [{\hat{\Delta\theta}}_{m1}^{\mathrm{c}},\ldots,{\hat{\Delta\theta}}_{mK}^{\mathrm{c}}]^{T} \in \mathbb{C}^{K \times 1}$. Correspondingly, we can adopt the updated joint angle and distance-based positioning similarity coefficient $\mathbf{\Xi}_{\mathcal{S}_{k'}^{\mathrm{h}}}^{\mathrm{c}}(\mathbf{\Theta}_k,\mathbf{\Psi}_k,\mathbf{\Theta}_{k'},\mathbf{\Psi}_{k'})$ to quantify the reward of the correction network $\boldsymbol{r}_{t}^{\mathrm{c}} = [r_{t,1}^{\mathrm{c}},\ldots,r_{t,M}^{\mathrm{c}}]$ with $r_{t,m}^{\mathrm{c}}$ at AP $m$, which can be expressed as (27), shown at the bottom of this page,
where $\mathcal{S}_{m}^{\mathrm{c,c}}$ denotes the evaluation subset of user positioning at AP $m$, which belongs to the mapping matrix of ${\mathcal{S}_{{k}}^{\mathrm{c,c}}}$ under the AOA-based supplementary correction network. Moreover, $\hat{\mathbf{\Psi}}_{k}^{m,\mathrm{c}}$ and $\hat{\mathbf{\Theta}}_{k}^{m,\mathrm{c}}$ represent the RSS value and AOA information extracted from the position of UE $k$ estimated by AP $m$, respectively, which both are determined by the initial position of AP $m$ and the actions output by the positioning and correction network, i.e., $x_{km}^{\mathrm{u,c}} = x_{m}^{\mathrm{a}} + \hat{d}_{mk}\mathrm{cos}(\hat{\theta}_{mk}^{\mathrm{p}} + {\hat{\Delta\theta}}_{mk}^{\mathrm{c}})$ and $y_{km}^{\mathrm{u,c}} = y_{m}^{\mathrm{a}} + \hat{d}_{mk}\mathrm{sin}(\hat{\theta}_{mk}^{\mathrm{p}}+{\hat{\Delta\theta}}_{mk}^{\mathrm{c}})$.

Similarly, the policy gradient of the AOA-based supplementary correction network for $\pi_{m}^{\mathrm{c}}$ can be modeled as
\begin{equation}
\setcounter{equation}{28}
\begin{aligned}
\nabla_{\theta_{\pi_m^\mathrm{c}}}J(\theta_{\pi_m^{\mathrm{c}}})=
{\mathbb{E}
\Big[\nabla_{\theta_{\pi_m^\mathrm{c}}}\pi_m^\mathrm{c}(\boldsymbol{a}_{t,m}^{\mathrm{c}})
Q_{\theta_{Q_{\pi_m^\mathrm{c}}}}(\mathbf{S}_{t}^{\mathrm{c}},\mathbf{A}_{t}^{\mathrm{c}})\Big]},
\label{eq27}
\end{aligned}
\end{equation}
where $\mathcal{D}^{\mathrm{c}}$ denote the experience extraction replay buffer of the supplementary correction network that stores experience $<\mathbf{S}_{t}^{\mathrm{c}}, \mathbf{A}_{t}^{\mathrm{c}}, \boldsymbol{r}_{t}^{\mathrm{c}}, \mathbf{S}_{t+1}^{\mathrm{c}}>$.

Additionally, the mean-squared Bellman error function \cite{[5]} of the current critic network $L({\theta_{Q_{\pi_m^\mathrm{c}}}})$ can be defined as
\begin{equation}
\setcounter{equation}{29}
\begin{split}
L({{\theta_{Q_{\pi_m^\mathrm{c}}}}}) = \mathbb{E}_{\mathbf{S}_{t}^{\mathrm{c}},\mathbf{A}_{t}^{\mathrm{c}}\sim \mathcal{D}^{\mathrm{c}}}\bigg[\Big(Q_{{\theta_{Q_{\pi_m^\mathrm{c}}}}}(\mathbf{S}_{t}^{\mathrm{c}},\mathbf{A}_{t}^{\mathrm{c}})-y_{t,m}^{\mathrm{c}}\Big)^2\bigg],
\label{eq28}
\end{split}
\end{equation}
where the action value $Q_{\theta_{Q_{\pi_m^\mathrm{c}}}}(\mathbf{S}_{t}^{\mathrm{c}},\mathbf{A}_{t}^{\mathrm{c}})$ is calculated by the current critic network ${{\theta_{Q_{\pi_m^\mathrm{c}}}}}$, and $y_{t,m}^{\mathrm{c}}=r_{t,m}^{\mathrm{c}}+\gamma^{\mathrm{c}}\Big(Q_{{\theta_{Q_{\pi_m^\mathrm{c,tar}}}}}(\mathbf{S}_{t}^{\mathrm{c,tar}},\mathbf{A}_{t}^{\mathrm{c,tar}})\Big)$ is the target value of agent $m$ with the target critic network value $Q_{{\theta_{Q_{\pi_m^\mathrm{c,tar}}}}}(\mathbf{S}_{t}^{\mathrm{c,tar}},\mathbf{A}_{t}^{\mathrm{c,tar}})$.

Furthermore, the soft update is carried out with $\tau^\mathrm{c} \ll 1$ to ensure that the target actor network ${\theta_{\pi_m^\mathrm{c,tar}}}$ and critic network ${{\theta_{Q_{\pi_m^\mathrm{c,tar}}}}}$ remains stable, which are given by
\begin{equation}
\setcounter{equation}{30}
\begin{split}
\left \{
\begin{array}{ll}
{\theta_{\pi_m^\mathrm{c,tar}}} \leftarrow \tau^\mathrm{c}{\theta_{\pi_m^\mathrm{c,tar}}}+(1-\tau^\mathrm{c}){\theta_{\pi_m^\mathrm{c}}},\\
{{\theta_{Q_{\pi_m^\mathrm{c,tar}}}}} \leftarrow \tau^\mathrm{c}{{\theta_{Q_{\pi_m^\mathrm{c,tar}}}}} + (1-\tau^\mathrm{c}){{\theta_{Q_{\pi_m^\mathrm{c}}}}}.
\end{array}
\right.
\label{eq29}
\end{split}
\end{equation}
\subsection{User Positioning Estimation}
To effectively estimate the position information of all UEs, we propose combining the action outputs by all APs through a joint positioning and correction network based on AOA and RSS information to achieve user positioning. As shown in Fig. 5, conventional estimation schemes, such as KNN-based and WKNN-based \cite{[8],[9]} eliminate the involvement of distantly located APs in user positioning through RSS information. However, they often overlook on AOA information, which may lead to the exclusion of APs that having strong correlations.
Additionally, the approach of selecting a fixed number of APs to participate in user positioning, there may be a weak correlation between the APs involved. Therefore, we combine RSS and AOA information to design similarity thresholds and select a variable number of APs to participate in user positioning, thereby improving positioning performance, termed Co-WKNN-based estimation scheme. Compared with the conventional estimation schemes, the positioning similarity in our proposed Co-WKNN-based estimation scheme under joint AOA and RSS information can better converge to a single tip, which can more accurately match the actual physical position $(x_k,y_k)$ with the estimated physical position $(\hat{x}_{k},\hat{y}_{k})$ of UE $k$, thereby enhancing positioning accuracy.

Then, the estimated position of UE $k$ can be represented as
\begin{equation}
\setcounter{equation}{31}
\begin{split}
\left \{
\begin{array}{ll}
\hat{x}_{k}  = \sum_{m \in \mathcal{M}_k^{\mathrm{CoW}}}w_{mk}^{\mathrm{CoW}}\Big(x_{m}^{\mathrm{a}} + \hat{d}_{mk}\mathrm{cos}(\hat{\theta}_{mk}^{\mathrm{p}} + {\hat{\Delta\theta}}_{mk}^{\mathrm{c}})\Big),\\
\hat{y}_{k} = \sum_{m \in \mathcal{M}_k^{\mathrm{CoW}}}w_{mk}^{\mathrm{CoW}}\Big(y_{m}^{\mathrm{a}} + \hat{d}_{mk}\mathrm{sin}(\hat{\theta}_{mk}^{\mathrm{p}} + {\hat{\Delta\theta}}_{mk}^{\mathrm{c}})\Big),
\end{array}
\right.
\label{eq30}
\end{split}
\end{equation}
where $w_{mk}^{\mathrm{CoW}}$ is the weight coefficient of the $m$-th AP selected under the Co-WKNN-based estimation scheme,
and $\mathcal{M}_k^{\mathrm{CoW}}$ denotes the set of APs selected for user positioning based on the joint information ${\mathbf{{\Xi}}_{\mathcal{S}_{{k}}^{\mathrm{c}}}^{\mathrm{j}}(\mathbf{\Psi}_k,\mathbf{\Theta}_k,\hat{\mathbf{\Psi}}_{k}^m,\hat{\mathbf{\Theta}}_{k}^m)}$, e.g., (32), shown at the bottom of this page, where $\mathbf{\Xi}_\mathrm{thres}$ is the joint distance and angle-based similarity threshold, satisfying $|\mathcal{M}_k^{\mathrm{CoW}}| = M_{\mathrm{c}}$ with the number of selected APs $M_{\mathrm{c}}$ and
\begin{equation}
\setcounter{equation}{33}
\begin{split}
w_{mk}^{\mathrm{CoW}}=\frac{\mathbf{{\Xi}}_{\mathcal{S}_{{k}}^{\mathrm{c}}}^{\mathrm{j}}(\mathbf{\Psi}_k,\mathbf{\Theta}_k,\hat{\mathbf{\Psi}}_{k}^m,\hat{\mathbf{\Theta}}_{k}^m)}{\sum_{l \in \mathcal{M}_k^{\mathrm{CoW}}}\mathbf{{\Xi}}_{\mathcal{S}_{{k}}^{\mathrm{c}}}^{\mathrm{j}}(\mathbf{\Psi}_k,\mathbf{\Theta}_k,\hat{\mathbf{\Psi}}_{k}^l,\hat{\mathbf{\Theta}}_{k}^l)}.
\label{eq32}
\end{split}
\end{equation}

\begin{table}[t]
\centering
    \fontsize{9}{8.45}\selectfont
    \caption{The Model Structure and Experimental Details.}
    \label{paper}
    \begin{tabular}{ccc}
    \toprule
    \bf Parameters &  \bf Size \\
    \midrule
    1st hidden layer & 128, Leaky Relu (0.01)\\
    2nd hidden layer & 64, Leaky Relu (0.01) \\
    Hidden layer (GNN) & 128, Relu \\
    Number of layers (GNN) & 5 \\
    Discounted factor $\gamma^{\mathrm{p}}$ and $\gamma^{\mathrm{c}}$ & 0.99 \\
    Experience pool size $\mathcal{D}^{\mathrm{p}}$ and $\mathcal{D}^{\mathrm{c}}$ & 64 and 512 \\
    Soft update rate $\tau^{\mathrm{p}}$ and $\tau^{\mathrm{c}}$ & 0.01 and 0.01 \\
    Random episodes & 2000\\
    \bottomrule
    \end{tabular}
\end{table}
Correspondingly, based on the estimated position $(\hat{x}_{k},\hat{y}_{k})$ of UE $k$, $\forall k \in \mathcal{K}$, obtained from the MARL network, we can adopt RMSE $e_\text{RMSE}$ to evaluate positioning accuracy.
\section{Numerical Results}
In this section, we provide some numerical results to evaluate the performance of the JPC-MADDPG algorithm under the Co-WKNN-based estimation scheme. We consider a scenario where $M$ APs and $K$ UEs are uniformly distributed in a $0.1 \times 0.1\,\mathrm{km^2}$ area, exploiting a wrap-around scheme \cite{[6]}. This setting includes a blend of complex indoor and small outdoor environments, such as indoor smart factories, outdoor courts, etc. Moreover, each coherence block consists of $\tau_c = 200$ and $\tau_p = K$ channel uses are reserved for uplink pilot transmission corresponding to a coherence bandwidth of 200 kHz and a coherence time of 1 ms. Also, we investigate communication with $B = 20$ MHz bandwidth and $\sigma^2=-94$ dBm noise power, satisfying $\sigma^2 = -174 + 10\text{\rm log}_{10}(B)+ \sigma^2_{\text{\rm figure}}$ with a noise figure $\sigma^2_{\text{\rm figure}}=7$ dB \cite{[6],[24]}. We set the height difference between the APs and UEs and the carrier frequency as 10 m and 10 GHz, respectively. The Rician $\kappa$-factor can be denoted as $\kappa_{mk}=10^{1.3-0.003 d_{mk}}$ with the distance $d_{mk}$ between AP $m$ and UE $k$ for the LoS condition and $\kappa_{mk}=0$ for the NLoS condition. Furthermore, phase shift and antenna spacing can be denoted as $\varkappa_{mk}=e^{j\mu_{mk}}$ with $\mu_{mk} \sim \mathcal{U}(0,2\pi)$ and $\Delta = \lambda/2$. Meanwhile, we utilize a typical three-slope propagation model to compute coefficient $\beta_{mk}$ between AP $m$ and UE $k$ \cite{[3]}.

Next, the performance of the proposed positioning algorithm is compared with the following benchmark schemes: (1) Learning-based schemes: multidimensional graph neural network (MDGNN)-aided positioning algorithms \cite{[39],[42]}, which can achieve performance similar to conventional centralized schemes but with reduced computational complexity, including one-dimensional (1D) MDGNN and two-dimensional (2D) MDGNN. Specifically, for MDGNN (1D) schemes, permutations involve either AP or UE positions, while MDGNN (2D) schemes, they effectively permute the joint order of AP and UE, thereby achieving better performance by fully utilizing permutation properties, but also increasing the required computational complexity. (2) Non-learning-based schemes: fingerprint positioning algorithms \cite{[9],[10],[11]}.
For baseline comparisons of computational complexity and positioning performance, we select the number of reference points under the fingerprint positioning algorithm as $N_\eta = 1600$ for the lower bound and as $N_\eta = 40000$ for the upper bound (e.g., the reference point spacing $\eta = 2.5$ and $\eta = 0.5$ m), respectively \cite{[7],[9],[10]}. Note that the selection of reference points and the collection of fingerprint datasets are preset and do not affect actual scenarios \cite{[7],[9]}. All experiments are conducted using Python 3.7 with an Nvidia GeForce GTX 3060 (6 GB) Graphics Processing Unit, and the detailed parameters are summarized in Table \uppercase\expandafter{\romannumeral2}.

\begin{table}[t]
    \caption{Comparison of Computational Complexity.}
	\label{tab2}
	\centering
	\footnotesize
	\renewcommand{\arraystretch}{1.25}
    \begin{tabular}{ll}
    \hline\hline
    \textbf{\makecell{Parameters}} &\textbf{\makecell{Computational Complexity}} \\ \hline\hline
    \multicolumn{2}{l}{\emph{\textbf{Fingerprint Positioning + WKNN}}} \\ \hline\hline
    \emph{\makecell{RSS}}  & \makecell{$\mathcal{O}(MM_{\mathrm{RP}}^2K)$} \\ \hline
    \emph{\makecell{AOA / JAR}}  & \makecell{$\mathcal{O}(MM_{\mathrm{RP}}^2KN)$} \\ \hline
    \multicolumn{2}{l}{\emph{\textbf{Multidimensional GNN-aided Positioning + Permutation Prior}}} \\ \hline\hline
    \emph{\makecell{MDGNN (1D)}}  & \makecell{$\mathcal{O}(MK\sum_{l=1}^{\mathrm{La-1}}C_l^{\mathrm{1D}}C_{l+1}^{\mathrm{1D}})$} \\ \hline
    \emph{\makecell{MDGNN (2D)}}  & \makecell{$\mathcal{O}(MK\sum_{l=1}^{\mathrm{La-1}}C_l^{\mathrm{2D}}C_{l+1}^{\mathrm{2D}})$} \\ \hline
    \multicolumn{2}{l}{\emph{\textbf{Joint Positioning and Correction MADDPG + Co-WKNN}}} \\ \hline\hline
    \emph{\makecell{RSS}} & \makecell{$\mathcal{O}(MKd_{a}\sum_{a=1}^{A}Q_a^{2}+M^2Kd_a\sum_{c=1}^{C}Q_c^{2})$}  \\ \hline
    \emph{\makecell{AOA / JAR}} & \makecell{$\mathcal{O}(MKNd_{a}\sum_{a=1}^{A}Q_a^{2}+M^2KNd_a\sum_{c=1}^{C}Q_c^{2})$} \\ \hline\hline
\end{tabular}
\end{table}
Additionally, we compare the computational complexity of different positioning algorithms, as shown in Table \uppercase\expandafter{\romannumeral3}.
For the proposed JPC-MADDPG algorithm, the computational complexity under an RSS-based scheme is $\mathcal{O}(MKd_{a}\sum_{a=1}^{A}Q_a^{2}+M^2Kd_a\sum_{c=1}^{C}Q_c^{2})$, where $A$ and $C$ are the number of hidden layers for implementing the actor and critic networks with the dimension of the output action $d_{a}^\mathrm{p}$ and $d_{a}^\mathrm{p}$, satisfying $d_{a}=d_{a}^\mathrm{p} + d_{a}^\mathrm{c}$, and $Q_a$ denotes the output size of the $a$-th layer or the input size of the next layer.
Compared with RSS-based scheme, the computational complexity under the JAR-based scheme is mainly determined by small-scale fading information proportional to the number of antennas per AP, which can be represented as $\mathcal{O}(MKNd_{a}\sum_{a=1}^{A}Q_a^{2}+M^2KNd_a\sum_{c=1}^{C}Q_c^{2})$.
For fingerprint positioning algorithms, the computational complexity under RSS-based scheme is $\mathcal{O}(MM_{\mathrm{RP}}^2K)$, which is proportional to the number of reference points $M_{\mathrm{RP}}$. Similarly, the computational complexity under the JAR-based scheme is $\mathcal{O}(MM_{\mathrm{RP}}^2KN)$, which is proportional to the number of antennas per AP.
Moreover, for MDGNN-aided positioning algorithms, the computational complexity under its one-dimensional and two-dimensional frameworks is $\mathcal{O}(MK\sum_{l=1}^{\mathrm{La}}C_l^{\mathrm{1D}}C_{l+1}^{\mathrm{1D}})$ and $\mathcal{O}(MK\sum_{l=1}^{\mathrm{La}}C_l^{\mathrm{2D}}C_{l+1}^{\mathrm{2D}})$, respectively, where $C_l^{\mathrm{1D}}$ ($C_l^{\mathrm{2D}}$) and $C_{l+1}^{\mathrm{1D}}$ ($C_{l+1}^{\mathrm{2D}}$) are the inputs and outputs of each layer of the network. $\mathrm{La}$ represents the number of network layers in MDGNN.
Moreover, for a one-dimensional framework, the first input layer and the last output layer satisfy $C_1^{\mathrm{1D}}=2ML$ and $C_{\mathrm{La}}^{\mathrm{1D}}=2L$ with $2C_l^{\mathrm{1D}}C_{l+1}^{\mathrm{1D}}$ trainable parameters, while for a two-dimensional framework, the first input layer and the last output layer satisfy $C_1^{\mathrm{2D}}=2M$ and $C_{\mathrm{La}}^{\mathrm{2D}}=2$ with $3C_l^{\mathrm{2D}}C_{l+1}^{\mathrm{2D}}$ trainable parameters. Note that the adopted MDGNN framework belongs to an unsupervised model, which has the same input data as the proposed algorithm, both of which are observed information \cite{[39]}.
It is clear that the proposed JPC-MADDPG algorithm can significantly reduce computational complexity and network overhead.
This is achieved by eliminating the establishment and search stages of the fingerprint database, thereby enhancing the feasibility of the positioning scheme in cell-free mMIMO systems.\\
\begin{figure}[t]
	\centering
	\begin{minipage}[t]{0.48\textwidth}
	\centering
    \includegraphics[scale=0.54]{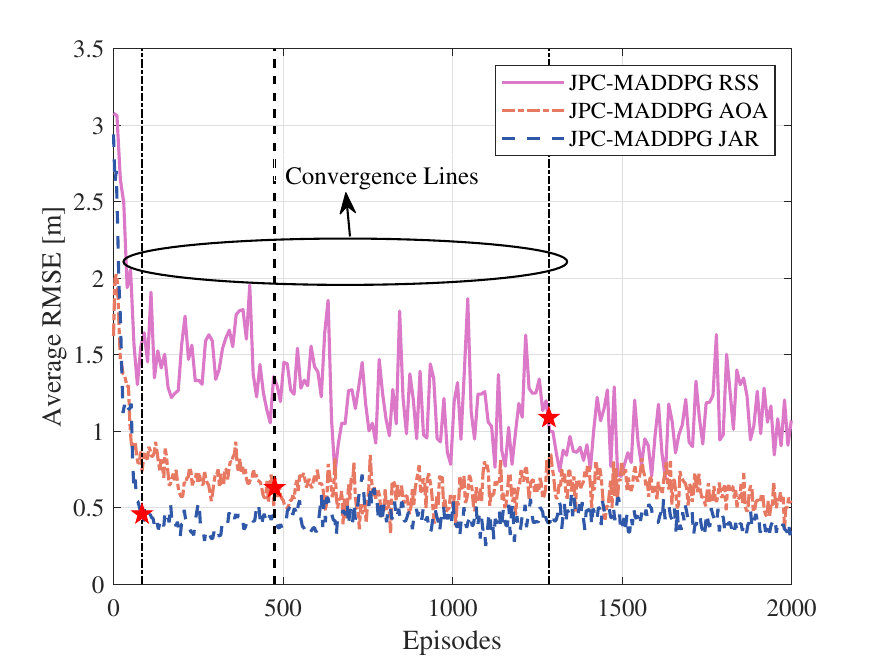}
    \caption{Convergence rate over different positioning similarity schemes with $M=36$, $K=9$, $N=8$, $L=6$, $\tau_p=K$, and $\kappa_{mk} = 0$.}
	\end{minipage}
    \hfill%
	\begin{minipage}[t]{0.48\textwidth}
	\centering
    \includegraphics[scale=0.54]{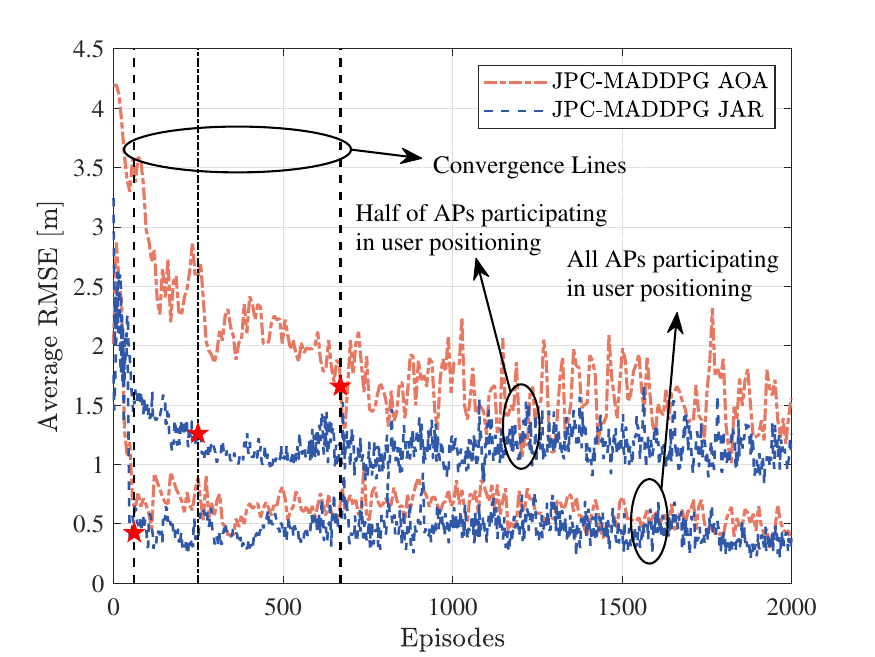}
    \caption{Convergence rate over different numbers of APs participating in positioning with $M=36$, $K=6$, $N=8$, $L=6$, $\tau_p=K$, and $\kappa_{mk} = 0$.}
	\end{minipage}
\end{figure}
\begin{figure}[t]
    \begin{minipage}[t]{0.48\textwidth}
	\centering
    \includegraphics[scale=0.54]{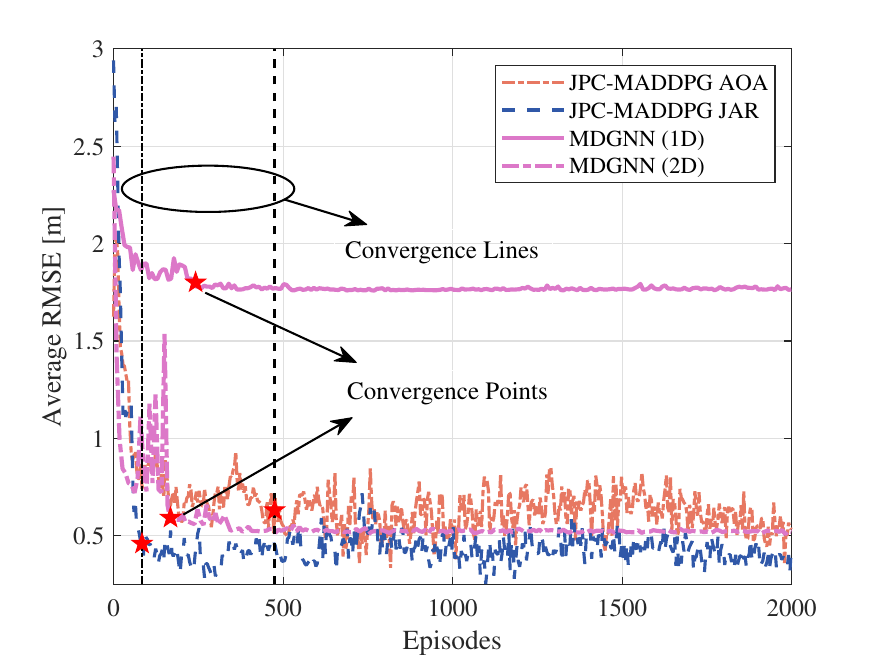}
    \caption{Convergence rate over different positioning frameworks with $M=36$, $K=9$, $N=8$, $L=6$, $\tau_p=K$, and $\kappa_{mk} = 0$.}
	\end{minipage}
    \hfill%
    \begin{minipage}[t]{0.48\textwidth}
    \centering
    \includegraphics[scale=0.54]{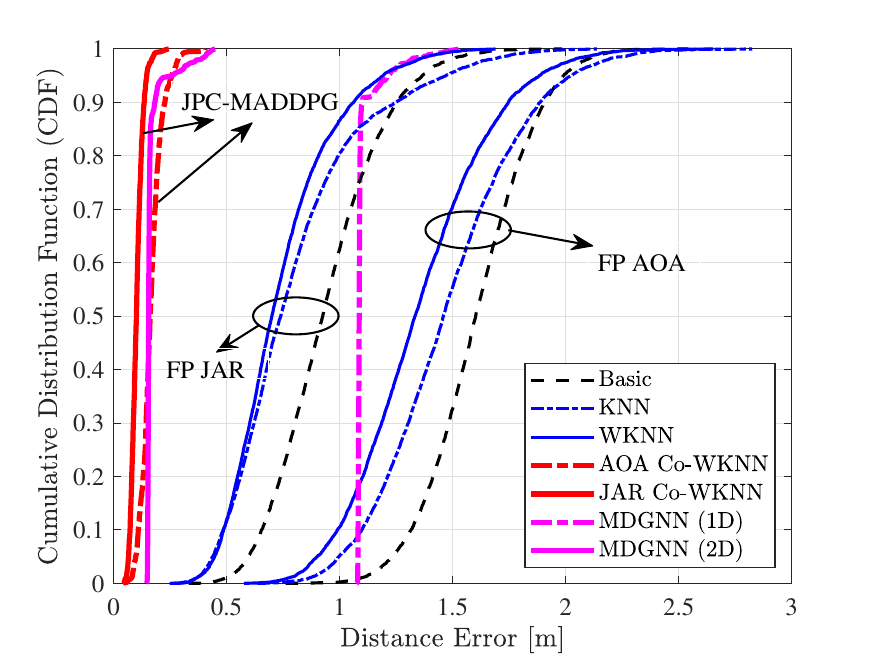}
    \caption{CDF of the average RMSE against different estimation schemes with $M=36$, $K=9$, $N=16$, $L=6$, $\tau_p=K$, and $\kappa_{mk} = 0$.}
	\end{minipage}
\end{figure}
\begin{figure}[t]
	\begin{minipage}[t]{0.48\textwidth}
	\centering
    \includegraphics[scale=0.54]{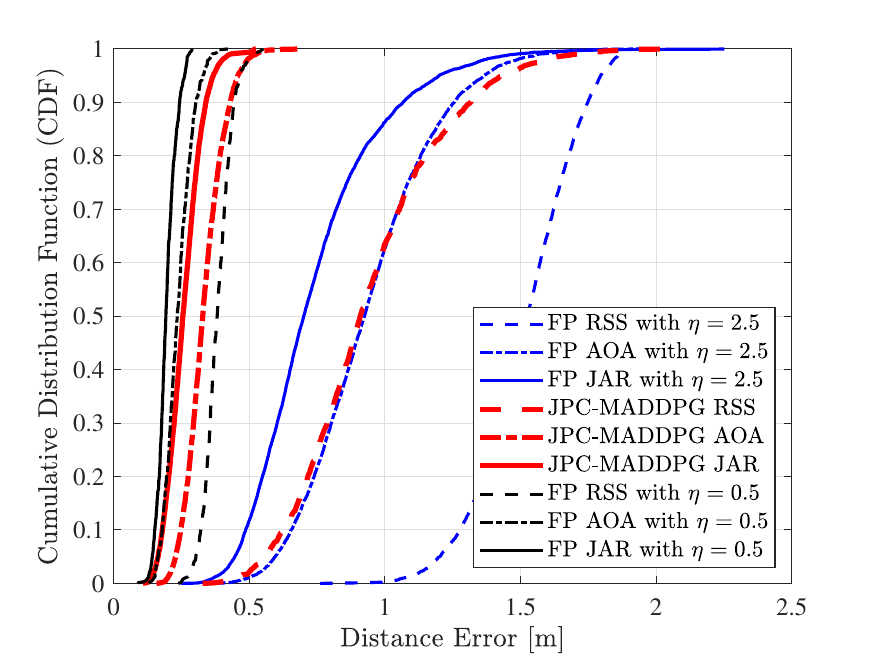}
    \caption{CDF of the average RMSE against different positioning similarity schemes with $M=36$, $K=9$, $N=8$, $L=6$, $\tau_p=K$, and $\kappa_{mk} = 0$.}
	\end{minipage}
    \hfill%
	\begin{minipage}[t]{0.48\textwidth}
	\centering
    \includegraphics[scale=0.54]{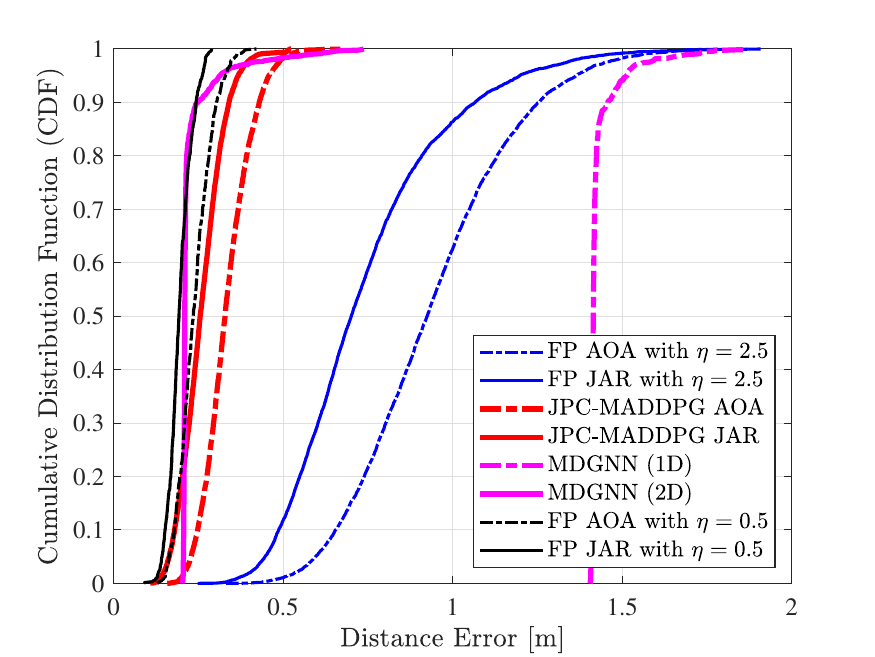}
    \caption{CDF of the average RMSE against different positioning frameworks with $M=36$, $K=9$, $N=8$, $L=6$, $\tau_p=K$, and $\kappa_{mk} = 0$.}
	\end{minipage}
\end{figure}
\begin{figure}[t]
	\centering
	\begin{minipage}[t]{0.48\textwidth}
	\centering
    \includegraphics[scale=0.54]{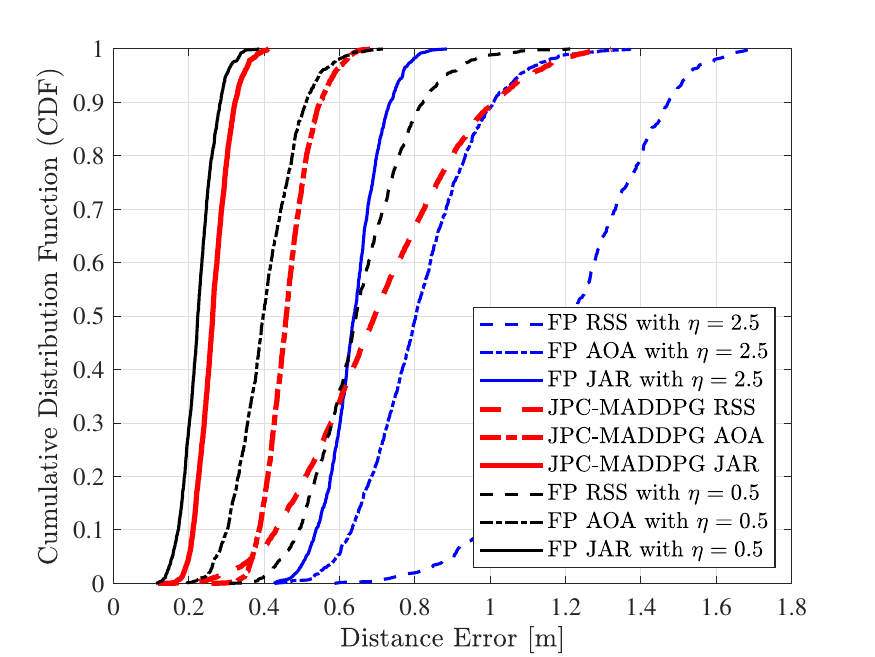}
    \caption{CDF of the average RMSE against different positioning similarity schemes with $M=36$, $K=9$, $N=8$, $L=6$, $\tau_p=K$, and $\kappa_{mk} \neq 0$.}
	\end{minipage}
    \hfill%
	\begin{minipage}[t]{0.48\textwidth}
	\centering
    \includegraphics[scale=0.54]{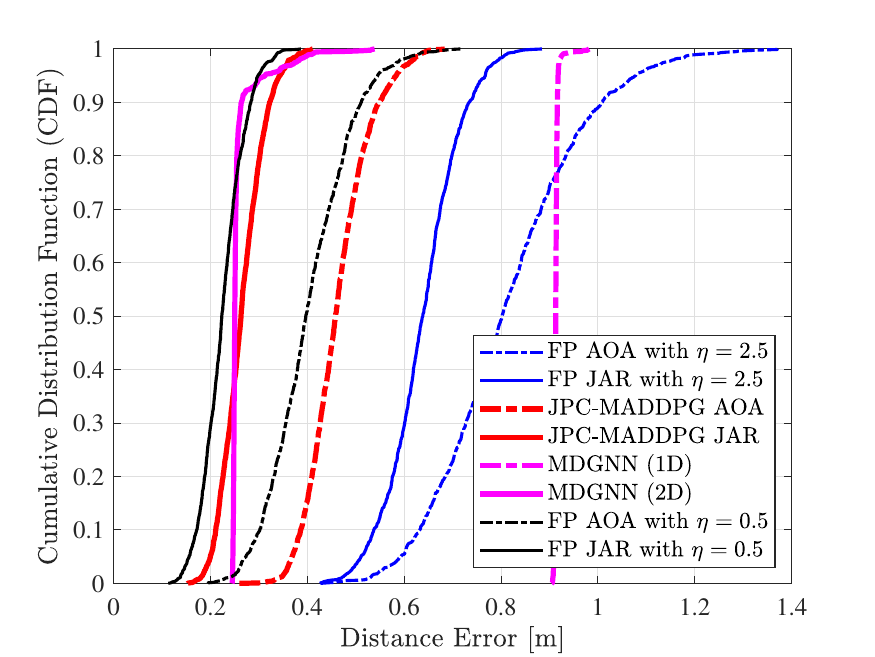}
    \caption{CDF of the average RMSE against different positioning frameworks with $M=36$, $K=9$, $N=8$, $L=6$, $\tau_p=K$, and $\kappa_{mk} \neq 0$.}
	\end{minipage}
\end{figure}
\begin{figure}[t]
	\centering
	\begin{minipage}[t]{0.48\textwidth}
	\centering
    \includegraphics[scale=0.54]{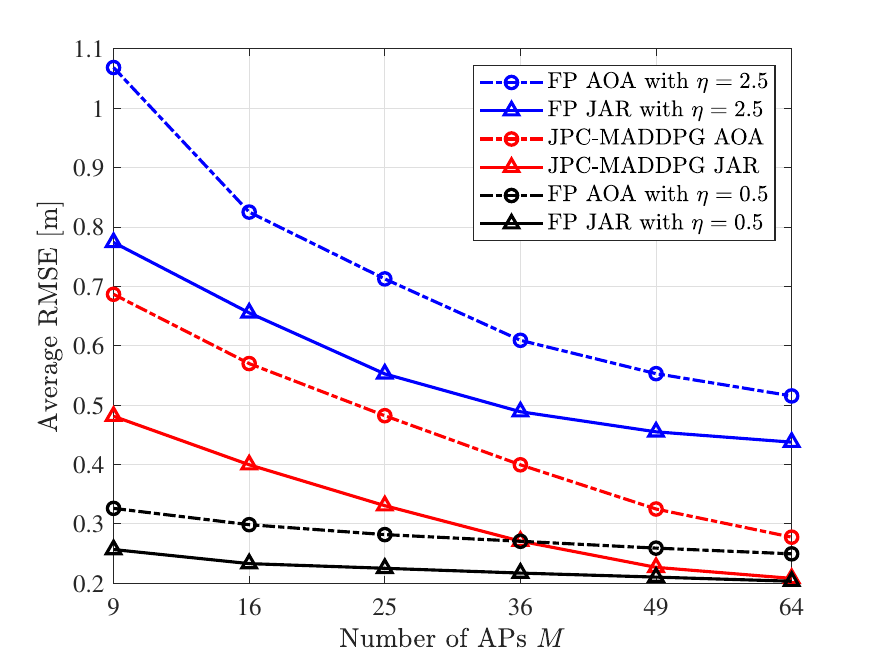}
    \caption{The average RMSE versus the number of APs with $K=6$, $N=8$, $L=6$, $\tau_p=K$, and $\kappa_{mk} = 0$.}
	\end{minipage}
    \hfill%
	\begin{minipage}[t]{0.48\textwidth}
	\centering
    \includegraphics[scale=0.54]{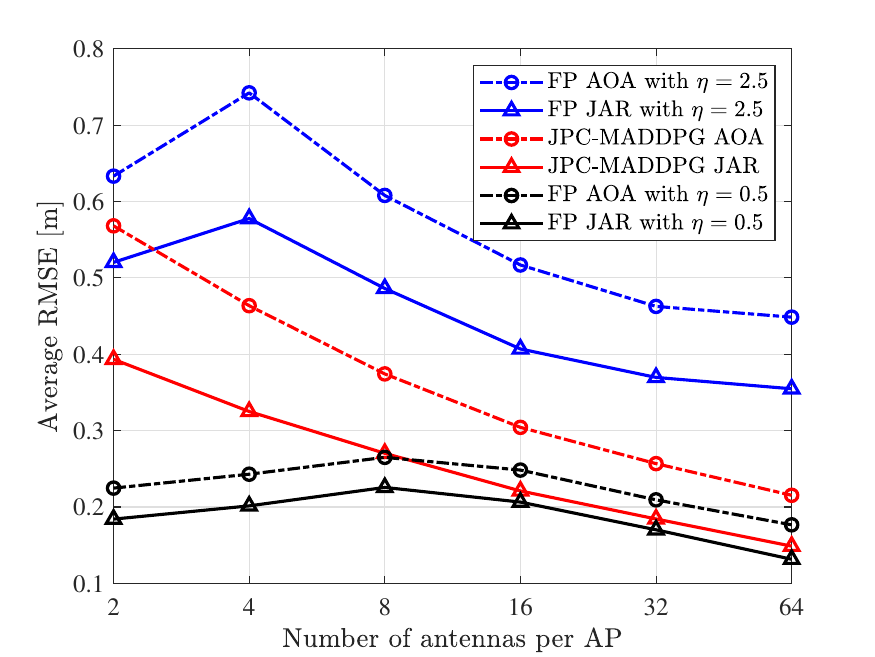}
    \caption{The average RMSE versus the number of antennas per AP with $M=36$, $K=9$, $L=6$, $\tau_p=K$, and $\kappa_{mk} = 0$.}
	\end{minipage}
\end{figure}
Fig. 6 shows the convergence rate over different positioning similarity schemes $\kappa_{mk}=0$ (only containing NLoS components), including RSS, AOA, and JAR. Compared with the conventional schemes, e.g., RSS-based and AOA-based, the novel JAR-based scheme achieves user positioning by fully extracting distance and angle information, resulting in improved convergence rates of 93.46\% and 82.32\%, respectively. This indicates that combining distance and angle information can effectively improve the convergence rate. On the other hand, Fig. 7 shows the impact of different numbers of APs participating in positioning evaluation on the convergence rate $\kappa_{mk}=0$ (only containing NLoS components), e.g., half of the APs and all APs participating in positioning evaluation. It is clear that increasing the number of APs participating in positioning evaluation can improve the convergence rate as the increased observation information from different physical locations facilitates more efficient user positioning. For example, compared with the case having half of the APs participating in positioning evaluation, all APs participate in positioning evaluation yields a 75.51\% improvement in convergence rate under the JAR-based scheme. Moreover, similar to Fig. 6, the convergence rate of the JAR-based scheme is always better than that of the AOA-based scheme, which is not affected by the number of APs participating in positioning evaluation, e.g., the convergence rate gap is 62.78\% under half of the APs participating in positioning evaluation. Moreover, Fig. 8 illustrates the convergence rate over different positioning frameworks $\kappa_{mk}=0$ (only containing NLoS components), including MADDPG-aided and MDGNN-aided. Compared to MDGNN (1D) and MDGNN (2D), our proposed JPC-MADDPG can achieve better balance convergence rate and positioning accuracy under the JAR-based scheme, by effectively combining angle and distance information, such as a 50.31\% improvement in convergence rate and a 17.18\% improvement in positioning accuracy compared to MDGNN (2D).

Fig. 9 compares the CDF of the average RMSE under various estimation schemes $\kappa_{mk}=0$ (only containing NLoS components), including basic, KNN-based, WKNN-based, Co-WKNN-based, and MDGNN-aided. It is evident that the proposed JPC-MADDPG algorithm and cooperative WKNN-based estimation significantly enhance positioning accuracy compared to conventional fingerprint algorithm. For example, under the JAR-based scheme, compared to fingerprint positioning with WKNN-based estimation, the positioning performance of the JPC-MADDPG algorithm with a cooperative WKNN-based estimation has shown a remarkable improvement of 97.49\%. This indicates that strategically eliminating APs with weaker correlations can effectively prevent the generation of incorrect positioning information, thereby achieving more precise user positioning.

Fig. 10 and Fig. 11 compare the CDF of the average RMSE between the proposed JPC-MADDPG algorithm and conventional algorithms under various positioning similarity schemes with $\kappa_{mk}=0$ (only containing NLoS components), including fingerprint positioning and MDGNN-aided positioning (1D and 2D). It is clear that the JAR-based scheme, which leverages comprehensive angle and distance information from all UEs, demonstrates superior accuracy in user positioning compared to conventional RSS-based and AOA-based schemes. For example, it demonstrates improvements of 31.16\% and 22.43\% in positioning accuracy under the JAR-based scheme, respectively. Moreover, compared to conventional positioning algorithms, our proposed JPC-MADDPG under the JAR-based scheme can better approximate the positioning performance of centralized positioning schemes with a faster convergence rate, including fingerprint positioning ($\eta = 0.5$) and MDGNN (2D), and is far superior to the positioning performance of fingerprint positioning ($\eta = 2.5$) and MDGNN (1D).

Moreover, Fig. 12 and Fig. 13 compares the CDF of the average RMSE between the proposed JPC-MADDPG algorithm and conventional algorithms under various positioning similarity schemes with $\kappa_{mk}\neq 0$ (containing LoS components and NLoS components). We can observe that the obtained results are similar to those of Fig. 10 and Fig. 11. Compared to RSS-based and AOA-based schemes, the proposed JPC-MADDPG algorithm under the JAR-based scheme can better adapt to the impact of various channel models by effectively utilizing distance and angle information, e.g., improving the performance of 85.22\% and 44.12\%, respectively, similar to the conclusions obtained under $\kappa_{mk}=0$ (only containing NLoS components). Furthermore, the proposed JPC-MADDPG algorithm can still approximate centralized fingerprint positioning ($\eta = 0.5$) and MDGNN (2D) under the JAR-based scheme. This indicates that the proposed algorithm exhibits good adaptability and scalability, effectively coping with the impact of multiple types of path loss models, including Rayleigh ($\kappa_{mk}= 0$) and Rician ($\kappa_{mk}\neq 0$) fading channels. Correspondingly, by combining the positioning accuracy in Fig. 11 and Fig. 13 with the computational complexity in Table \uppercase\expandafter{\romannumeral3}, we can clearly observe that the proposed JPC-MADDPG algorithm can achieve positioning performance approaching that of conventional fully centralized schemes with lower computational complexity. This indicates that cooperative positioning architecture is a critical element in striking a balance between two indicators.
\begin{figure}[t]
\centering
    \includegraphics[scale=0.54]{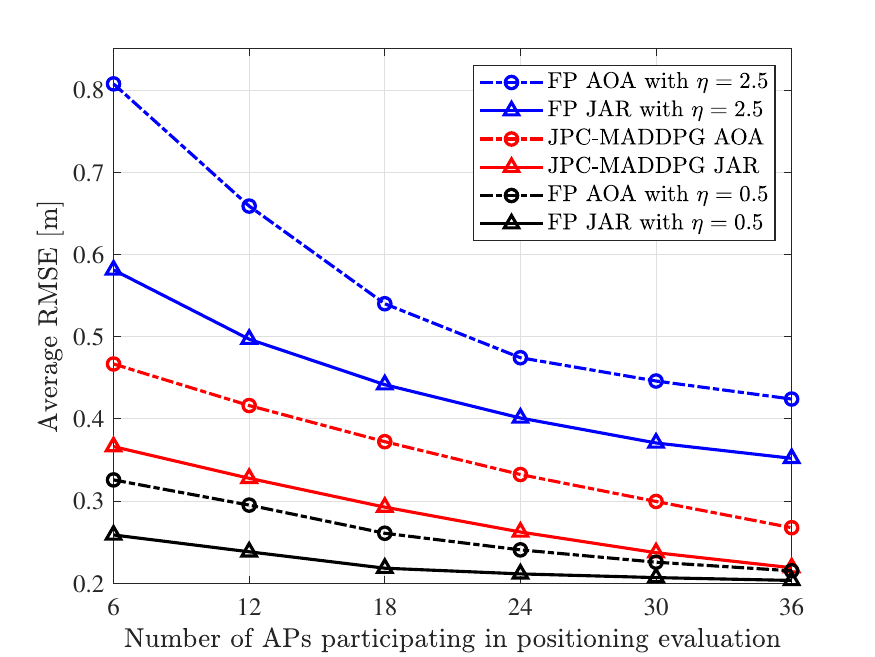}
    \caption{The average RMSE versus the number of APs participating in positioning with $M=36$, $K=9$, $N=8$, $L=9$, $\tau_p=K$, and $\kappa_{mk} = 0$.
    \label{fig12}}
\end{figure}
Fig. 14 and Fig. 15 show the average RMSE per UE as a function of the number of APs $M$ and antennas per AP $N$, respectively, with different user positioning schemes. It is clear that increasing the number of reference points can significantly reduce the estimation error under the fingerprint positioning algorithm, as it reduces the reference point spacing in user positioning, e.g., the positioning performance of fingerprint positioning at reference point spacing $\eta = 2.5$ exceeds 51.58\% of that at reference point spacing $\eta = 2.5$. Fig. 14 shows that as the number of APs increases, the positioning error gradually decreases, and the positioning performance of the proposed JPC-MADDPG gradually approaches that of the fingerprint positioning with $\eta = 0.5$. This indicates that increasing the number of APs can promote synergy between each other to obtain more accurate angle and distance information, thereby achieving user positioning.
Moreover, in Fig. 15, compared to the fingerprint positioning with $\eta = 2.5$ or $\eta = 0.5$, the proposed JPC-MADDPG adopts a cooperative WKNN-based estimation to filter the APs participating in user positioning, selecting highly correlated APs to better adapt to the increase in antenna numbers and reduce positioning errors. This reveals the importance of a cooperative estimation in improving scalability and generalization capabilities.

Fig. 16 shows the average RMSE versus the number of APs participating in positioning evaluation with different user positioning schemes. Consistent with the observations in Fig. 14, an increase in the number of APs participating in positioning evaluation reduces the positioning error. This is attributed to the fact that the angle and distance information of each UE can be better extracted by more participating APs. For example, as the number of APs participating in positioning evaluation increases from $6\,(L/6)$ to $36\,(L)$, the proposed JPC-MADDPG combining the cooperative WKNN estimation architecture yields a 42.67\% and 40.21\% improvement in positioning performance under AOA-based and JAR-based schemes, respectively. This indicates that the appropriate number of APs participating in positioning evaluation is crucial to strike an effective balance between communication overhead and positioning performance.
\section{Conclusion}
In this paper, we first adopted DFT operation to establish an angular domain channel model for cell-free mMIMO systems and proposed a joint distance and angle-based positioning similarity scheme within a cooperative architecture, selectively engaging highly correlated APs. Then, we proposed a novel MARL-based positioning scheme, which includes a preliminary positioning network, utilizing RSS information, and an auxiliary correction network, leveraging AOA information, to address user positioning problems.
The simulation results verified that the designed JPC-MADDPG adopting the Co-WKNN-based estimation can effectively enhance positioning accuracy while significantly reducing computational complexity.
In future work, we aim to extend the considered far-field region to the near-field region to investigate user positioning problems in mixed-field environments and explore the performance impact of mixed-field conditions on the effectiveness of our designed positioning scheme.
\bibliographystyle{IEEEtran}
\bibliography{IEEEabrv,Ref}
\end{document}